\documentclass[acmsmall]{acmart} 

\AtBeginDocument{%
  \providecommand\BibTeX{{%
    \normalfont B\kern-0.5em{\scshape i\kern-0.25em b}\kern-0.8em\TeX}}}

\setcopyright{rightsretained} 

\usepackage{enumitem}
\usepackage{multirow}
\usepackage{pifont} 
\newcommand{\cmark}{\ding{51}}%
\newcommand{\xmark}{\ding{55}}%

\begin{document}

\title[Sociotechnical Audits]{Sociotechnical Audits: Broadening the Algorithm Auditing Lens to Investigate Targeted Advertising}

\author{Michelle S. Lam}
\email{mlam4@cs.stanford.edu}
\orcid{0000-0002-3448-5961} 
\affiliation{%
  \institution{Stanford University}
  \city{Stanford}
  \state{California}
  \country{USA}
}

\author{Ayush Pandit}
\email{apandit@stanford.edu}
\orcid{0000-0003-2630-8356}
\affiliation{%
  \institution{Stanford University}
  \city{Stanford}
  \state{California}
  \country{USA}
}

\author{Colin H. Kalicki}
\email{ck9898@stanford.edu}
\orcid{0009-0002-6353-8492}
\affiliation{%
  \institution{Stanford University}
  \city{Stanford}
  \state{California}
  \country{USA}
}

\author{Rachit Gupta}
\email{rgupta432@gatech.edu}
\orcid{0000-0001-9705-9057}
\affiliation{%
  \institution{Georgia Institute of Technology}
  \city{Atlanta}
  \state{Georgia}
  \country{USA}
}

\author{Poonam Sahoo}
\email{pnsahoo@stanford.edu}
\orcid{0000-0002-9121-5378}
\affiliation{%
  \institution{Stanford University}
  \city{Stanford}
  \state{California}
  \country{USA}
}

\author{Dana\"e Metaxa}
\email{metaxa@seas.upenn.edu}
\orcid{0000-0001-9359-6090}
\affiliation{%
  \institution{University of Pennsylvania}
  \city{Philadelphia}
  \state{Pennsylvania}
  \country{USA}
}

\renewcommand{\shortauthors}{Michelle S. Lam et al.}

\begin{abstract}
    Algorithm audits are powerful tools for studying black-box systems without direct knowledge of their inner workings. While very effective in examining technical components, the method stops short of a sociotechnical frame, which would also consider users themselves as an integral and dynamic part of the system. 
Addressing this limitation, we propose the concept of \textit{sociotechnical auditing}: auditing methods that evaluate algorithmic systems at the sociotechnical level, focusing on the interplay between algorithms and users as each impacts the other. Just as algorithm audits probe an algorithm with varied inputs and observe outputs, a sociotechnical audit (STA) additionally probes users, exposing them to different algorithmic behavior and measuring their resulting attitudes and behaviors. 
As an example of this method, we develop \textit{Intervenr}, a platform for conducting browser-based, longitudinal sociotechnical audits with consenting, compensated participants. Intervenr investigates the algorithmic content users encounter online, and also coordinates systematic client-side interventions to understand how users change in response. 
As a case study, we deploy Intervenr in a two-week sociotechnical audit of online advertising ($N=244$) to investigate the central premise that personalized ad targeting is more effective on users. In the first week, we observe and collect all browser ads delivered to users, and in the second, we deploy an ablation-style intervention that disrupts normal targeting by randomly pairing participants and swapping all their ads.
We collect user-oriented metrics (self-reported ad interest and feeling of representation) and advertiser-oriented metrics (ad views, clicks, and recognition) throughout, along with a total of over $500,000$ ads. 
Our STA finds that targeted ads indeed perform better with users, but also that users begin to acclimate to different ads in only a week, casting doubt on the primacy of personalized ad targeting given the impact of repeated exposure. 
In comparison with other evaluation methods that only study technical components, or only experiment on users, sociotechnical audits evaluate sociotechnical systems through the interplay of their technical and human components. 

\end{abstract}


\acmJournal{PACMHCI}
\acmYear{2023} \acmVolume{7} \acmNumber{CSCW2} \acmArticle{360} \acmMonth{10} \acmPrice{}\acmDOI{10.1145/3610209}


\begin{teaserfigure}
  \includegraphics[width=\textwidth]{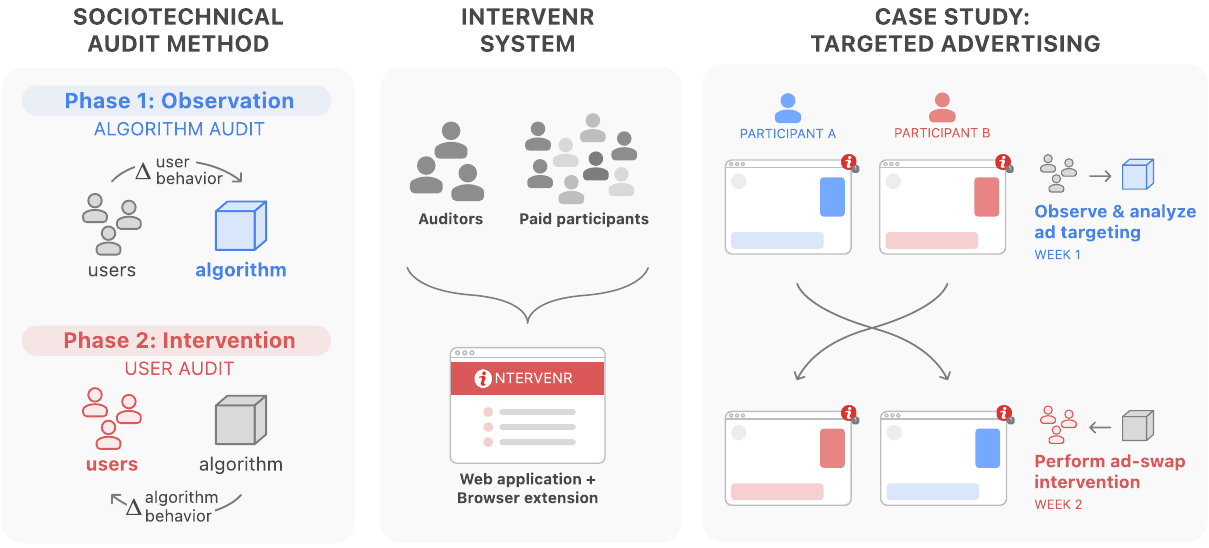}
  \caption{
  \textit{Method}---Sociotechnical audits evaluate algorithmic systems through a sociotechnical lens, evaluating the technical system and its impacts on users as both influence each other.
  \textit{Intervenr System}---We introduce a platform for deploying longitudinal, browser-based, user-centered sociotechnical audits with paid participants.
  \textit{Case Study}---We conduct a two-week sociotechnical audit of targeted advertising (N=244).\\
  }
  \label{fig:teaser}
\end{teaserfigure}

\maketitle

\section{Introduction}
The need to consider the human dimension of computational systems---how systems are used in practice, by whom, and to what end---is a central premise of research in Computer Supported Cooperative Work (CSCW). The idea of a \textit{sociotechnical system} is a prime example of this lens. 
While computer scientists commonly define a system in terms of its technical components, a sociotechnical frame expands the view of a system to include the critical interaction between technology and people \cite{cherns1976principles}. 
For example, when we consider an online advertisement platform as a sociotechnical system, we consider not just an algorithmic system that stores, ranks, and renders online ads, but also the end users who may interact with ads, be influenced by their content, and in turn influence how the algorithm functions.

Algorithm auditing has emerged over the last decade as a powerful and important tool for understanding black-box algorithmic systems and the content they output~\cite{sandvig2014auditing}. 
Algorithm audits are evaluations of existing algorithmic systems generally conducted by an external party, with the goal of identifying disparities with respect to a set of existing policies (often legal or ethical obligations). They do so by querying the system with different inputs and measuring its outputs to draw inferences about the system despite limited visibility into it~\cite{metaxa2021auditing}. Algorithm audits have proven especially effective for investigating pre-specified biases in the outputs of individual platforms; notable examples include audits of racial bias in algorithmic risk tools used in legal settings~\cite{propublica2016machine}, and gender and racial bias in facial recognition systems~\cite{buolamwini2018genderShades}. Today's algorithm audits are powerful tools for evaluating the technical components of a system and holding platforms accountable for biases at the algorithmic level.

Taking a sociotechnical view, however, there are limits to what an algorithm audit can uncover. While algorithm audits can reveal harmful patterns in a system's algorithmic output or responses to problematic user inputs, these audits are not designed to capture the substantial role of users in such systems. For example, an algorithm audit cannot capture how users interpret system outputs, how those outputs might modify user behaviors and beliefs to alter their use of the system, or how the system might change in response to changes in user behavior---especially over longer periods of time and at scale~\cite{matias2023humans, matias2023influencing}.

\subsubsection*{The Sociotechnical Audit Method} 
In this work, we argue that algorithm auditors seeking to understand the behavior of an algorithmic system would benefit from taking a sociotechnical approach that examines not just technical outcomes, but also the human outcomes that are so tightly coupled with them. 
Along those lines, we propose the concept of a \textit{sociotechnical audit} (or \textit{STA}), which mirrors the shift from studying purely technical systems to conceiving of some systems as sociotechnical and studying them accordingly. 
In addition to auditing the technical components of a system as with existing algorithm audit methods, a sociotechnical audit also evaluates the human portion of the system by conducting experiments that systematically intervene upon a user's experience with an algorithm as the two interact.
This method provides an opportunity to widen the lens that auditors use when identifying their object of study, shifting from the algorithm itself to the broader sociotechnical context in which that algorithm is situated. 

\subsubsection*{Intervenr: A System for Sociotechnical Auditing} 
It is challenging to instantiate this kind of audit: just as algorithm audits must gain an understanding of technical components by probing them with varied inputs and observing outputs~\cite{metaxa2021auditing}, sociotechnical audits must gain an understanding of human components by exposing users to varied algorithmic behavior and observing the impact on user attitudes and behaviors. And, as with all audits, auditors must do so without the ability to directly manipulate the algorithm or its users.
Most current work that falls within our definition of sociotechnical auditing bridges this gap by pairing algorithm auditing with separate controlled experiments on users (i.e., auditing ad content users see in-browser and then running separate experiments outside the browser to study responses to different ad content).
While valuable, we believe that these kinds of STAs can become much more impactful when they investigate the interplay of both the algorithm and its users as they interact in the real world. To audit the user side of a sociotechnical system at the same time as the algorithm, we can coordinate systematic client-side interface modifications---for example, auditing an ad targeting algorithm by collecting ad content in users' browsers, and simultaneously experimenting on them by altering the ads they see before the ads are delivered to them.

We develop a system called \textit{Intervenr} that allows researchers to conduct sociotechnical audits in the web browsers of consenting, compensated participants. Comprising a browser extension and web application, Intervenr is designed to perform sociotechnical audits in two phases. In the initial observational phase, Intervenr collects baseline observational data from a range of users to audit the technical component of the sociotechnical system. Then in the intervention phase, Intervenr enacts \textit{in situ} interventions on participants' everyday web browsing experience, emulating algorithmic modifications to audit the human component.

\subsubsection*{Case Study: Targeted Advertising} 
We demonstrate the value of sociotechnical auditing by deploying Intervenr on a case study of online advertising, investigating targeted ad content and its impact on users. 
To do so, we integrate an open-source ad blocker into the Intervenr system that identifies image and text advertisements loaded in users' browsers across all websites visited, and recruit 244 participants to use the system for a two-week audit study. 
Prior audits of targeted advertising have focused on the technical side of the equation, identifying problematic behaviors in ad targeting systems, such as user privacy leaks and biased ad distribution~\cite{faizullabhoy2018facebook, sapiezynski2022algorithms,kaplan2022impliedIdentityAds}. However, the impact of these systems on users remains notoriously opaque~\cite{schnadower2023behavioral}. This state of affairs is largely intentional on the part of ad intermediaries, whose control of the marketplace is strengthened by a lack of transparency~\cite{hwang2020subprime}. Given this opacity and ad targeting's reliance on invasive data collection and inference practices, questions remain regarding how targeted ad content impacts users over time, and whether its costs are justified---questions that require a sociotechnical approach to answer. Building on past audits of ad targeting, we use Intervenr to conduct an STA studying real users' behaviors and beliefs in the context of their targeted ads, and how those users respond to an alternative ad delivery algorithm.
To investigate the assumption that personalized targeting performs better with users, we design and deploy an ablation-style intervention: after collecting a baseline for each participant's ads, we randomly pair participants in the study and swap all their ads, disrupting normal targeting. This ablation allows us to measure the effect that targeted ads have on users and the extent to which user responses to ads are degraded after we break targeting. 

In the first week of our case study, we passively observe all ads delivered to participants. This traditional audit portion of our study allows us to measure canonical metrics like views and clicks, but also important dimensions at the locus of the user, like users' interest and feeling of representation as they relate to ad targeting. In the second week, we randomly pair participants, swapping each participant's ads with ads originally targeted to their partner. In addition to observing user behavior, we conduct participant surveys after each study phase that cover a subset of the ads collected; together, these produce both user-oriented metrics (ad interest and feeling of representation in ads) and advertiser-oriented metrics (ad views, clicks, and recognition).
Over the two-week study, we collect over 500,000 advertising images targeted to our study participants. Overall, we find that participants' own targeted ads outperform their swap partners' ads on all measures throughout the study, supporting the premise of targeted advertising. However, we also observe that swap partners' ads perform more highly with users at the close of the study (after only a week of exposure) than at the midpoint (before participants were exposed to their partners' ads). This is evidence that participants acclimate to their swap partners' ads, suggesting much of the efficacy of ad targeting may be driven by repeated exposure rather than the intrinsic superiority of targeting. 
While an algorithm audit could reveal whether today’s \textit{existing} targeting methods provide user benefit, our sociotechnical audit allows us to discover how that user benefit changes in response to \textit{alternative} algorithmic methods. In particular, this approach reveals that user sentiment toward ads may be more malleable than we expect, and casts doubt on the necessity of hyper-personalized and privacy-invasive targeting methods.

\subsubsection*{Contributions} 
In summary, our paper introduces three main contributions:
\begin{itemize}
    \item \textbf{The sociotechnical audit}. We introduce the concept of sociotechnical auditing: methods which extend algorithm auditing's focus on technical components to additionally audit the human components of a sociotechnical system.
    \item \textbf{The Intervenr platform}. We design and develop a platform for conducting sociotechnical audits. This system coordinates observation and \textit{in situ} interventions in participants' web browsers to emulate modified algorithm behavior and measure its impact on users.
    \item \textbf{An sociotechnical audit of targeted online advertising}. We use Intervenr to audit targeted in-browser advertising in a two-week sociotechnical audit ($N=244$). Using an ablation-style intervention that disrupts ad targeting, we investigate the assumption that targeted advertising performs better on users.
\end{itemize}

By conducting audits that conceive of algorithmic systems as sociotechnical and investigate both their technical and human components, we can form a richer understanding of these systems in practice. Sociotechnical audits can aid us in proposing and validating alternative algorithm designs with an awareness of their impact on users and society.

\section{Related Work}
\label{sec:relwork}

Sociotechnical auditing emerges from a combination of literatures on algorithm auditing and other experimental methods. 
In this section, we first cover algorithm auditing, which tends to comprise a technical audit, but not a user audit. Then, we describe several user-centered evaluation methods, which usually carry out user audits, but not technical audits. Throughout, we note exceptions that combine both perspectives, and which we would consider sociotechnical audits. When possible, we use examples of work on targeted advertising specifically to situate our case study amidst prior research in this space.

\subsection{Algorithm Auditing}
A unifying feature of auditing since its inception---including algorithm, sociotechnical, and even social science or financial audits---is the end goal of \textit{accountability}. An audit holds an entity accountable by conducting an evaluation with respect to a set of policies (often legal or ethical obligations)~\cite{flint1988philosophy}. Algorithm auditing is a particular auditing method commonly used to study technical systems. Inspired by audit studies in the social sciences, whose goals were to enforce non-discrimination law, algorithm audits are often used to uncover bias or discrimination~\cite{sandvig2014auditing}. These need not be the only policies studied; holding ad intermediaries accountable regarding their claims about the efficacy of targeted ads is one example. Notable algorithm audits have previously studied algorithmic systems in domains including employment~\cite{chen2018investigating, chen2015peeking}, housing \cite{edelman2017guests, asplund2020auditing}, web search~\cite{kay2015unequal, lai2019webSearch, metaxa2019searchMedia, robertson2018auditingPartisan}, healthcare~\cite{obermeyer2019dissecting}, and facial recognition~\cite{raji2019actionableAuditing, buolamwini2018genderShades}. Advertising is another popular domain for algorithm audits, as reviews of the space have found~\cite{bandy2021problematicMachineBehavior} and as references throughout this section will demonstrate. 

All sociotechnical audits must contain an algorithm audit to understand the technical aspect of the system, but most algorithm audits do not include an audit of users.
One reason for the absence of real users is that not all technical systems are viewed as sociotechnical---though we would encourage future auditors to consider that lens. For instance, Gender Shades, a high-profile audit of facial recognition algorithms' performance on faces of different gender presentations and skin tones, initially strictly considered the performance of the technical systems, not their production or use in practice~\cite{buolamwini2018genderShades}. However, followup work from the same authors found that disclosing biases to the responsible companies led them to build less biased products, which we would consider a sociotechnical audit of the process of engineering facial recognition systems. That followup included not only an audit of various software systems, but also an audit of how the people building those products acted differently in response to auditor pressure~\cite{raji2019actionableAuditing}.

Often, auditors choose to exclude real users from an audit to achieve greater experimental control and thus a stronger technical audit. Much work in this space has collected data using sock puppet accounts and other researcher-fabricated data collection strategies. One of the first such studies, from 2015, addressed bias in Google's ad targeting. Researchers built a tool called AdFisher to run experiments studying the impact of Ad Settings and user behavior on the ads received by sock puppets~\cite{datta2014automated}. The use of sock puppets allowed researchers to evaluate the algorithm in the context of identical user behavior. AdFisher revealed some biases in Google's ad targeting, for example that accounts registered as male received more ads for high-paying jobs. Later work used sock puppets to study gender and racial biases in housing ads with similar results~\cite{asplund2020auditing}. Recent research isolated the impact of perceived demographics in ad images on ad delivery algorithms by carefully controlling ad imagery through synthetic image generation~\cite{kaplan2022impliedIdentityAds}.
 
In other cases, auditors may choose to audit only the technical aspect of a system when the technical results alone constitute serious user harm, or when user impact is infeasible or unethical to study. 
Algorithm audits have found biases against women in resume ranking sites~\cite{chen2018investigating}, and that Google Search queries for Black-sounding names were more likely to result in ads suggestive of an arrest record than searches for white-sounding names~\cite{sweeney2013discrimination}. Such results have obviously problematic implications, even before proceeding to study user impacts in practice. 
Other audits demonstrated that Facebook's advertiser controls allowed discriminatory targeting of employment and housing ads~\cite{ProPublica2016facebook, speicher2018potential} and created gender-, racially-, and politically-biased audiences in response to advertiser content~\cite{ali2019discrimination, lambrecht2019algorithmic, ali2021political, sapiezynski2022algorithms}. Most work in this space has focused on Facebook's ad systems, but researchers have found similar results auditing the ad targeting systems of other platforms like Google, LinkedIn, and others~\cite{venkatadri2020potential, imana2021auditing, chang2021targeted}. These audits did not evaluate how the tools were used in practice by advertisers or the downstream impact of these ads on people. Nevertheless, enabling such discrimination violates U.S. anti-discrimination law, and led to a lawsuit brought by the U.S. Department of Housing and Urban Development, not to mention public outcry~\cite{npr2019housing}. 
Finally, in some cases, a full sociotechnical audit may be impossible: investigative journalists auditing the use of an algorithm in bail and sentencing decisions could not have experimented on the human users of the algorithm (judges) by having them make rulings using a range of different algorithms for obvious ethical and logistical reasons~\cite{propublica2016machine}.

\subsubsection{Crowdsourced Audits}
Recently, more audits have started to include end user perspectives, for example by collecting data directly from users (a method known as crowdsourced or collaborative auditing)~\cite{sandvig2014auditing, lurie2021searching, mustafaraj2020voterCentered}, or by empowering users themselves to run their own audits~\cite{lam2022end, shen_everyday_alg_audit}. Such audits generally collect real user data, for example by running search engine queries and collecting the results on users' machines~\cite{robertson2018auditingPartisan} or collecting the ads shown to users in their own Facebook feeds~\cite{nyu2021adobserver}. 
Despite the presence of real users, carrying out a user audit using these techniques is challenging, since auditors typically lack access to modify the system. In a recent example, researchers took a crowdsourced auditing approach to measure ad targeting at scale; they both audited the technical system and investigated user perceptions of targeting, but they did not have access to modify the technical system and measure how user perceptions might change~\cite{zeng2022Targeting}.
Mozilla has been active in this space with two projects, RegretsReporter and Rally. RegretsReporter crowdsources problematic algorithmic content from YouTube~\cite{mozilla2021regrets}. While it considers user experiences, RegretsReporter similarly cannot modify algorithmic behavior and thus does not carry out a user audit. On the other hand, Rally collects participants' browsing data, allows approved researchers to access aggregate data, and can also run lightweight in-browser experiments that modify user experiences~\cite{mozilla2021rally}. 
Given their dual capabilities to audit algorithms and audit users, we view Rally experiments as instances of sociotechical audits.

\subsection{User-Centered Evaluation Methods}
As audit scholars outside the technical setting have noted, auditing practices may share methodological similarities with other evaluation practices, though the positionality and goals of auditors generally differ from those of other evaluators~\cite{pollitt1996performance}. Below we examine several other user-centered evaluation methods and their relationship to sociotechnical auditing. 

\subsubsection{Descriptive Methods}
Descriptive methods like surveys can provide valuable insight into the users of a sociotechnical system, and they can sometimes act as user audits when they probe users with various inputs to observe their responses. These methods have been valuable in studying targeted ads, including research on user beliefs and behaviors around ad blockers~\cite{mathur2018characterizing}, as well as research characterizing and taxonomizing what users perceive as problematic ads~\cite{zeng2021problematic}. Other descriptive methods include interviews with users. For example, in a study that surfaced to participants specific targeted ads and the inferences on which they were based, researchers found that some inferences were considered ``creepy'' and that others caused people to feel disillusioned with the power of the algorithm~\cite{eslami2018adverts}.

\subsubsection{Randomized Controlled Trials and Field Experiments}
Randomized Controlled Trials (RCTs) are a method used in a wide range of disciplines to run experiments. When researchers seek to draw causal claims in the face of factors they cannot directly control, RCTs provide control through randomization. They are an effective and frequently used strategy to understand users of sociotechnical systems, and can be considered audits of user behavior. 
For example, in the ads domain, past work investigated the efficacy of affiliate advertisement disclosures on social media platforms by conducting randomized controlled experiments that presented differently-worded disclosures to crowdworkers~\cite{mathur2018endorsements}.
Researchers have also compared participant responses carefully-matched targeted ads and search ads as well as random products to investigate whether behavioral advertising indeed provides value to consumers~\cite{schnadower2023behavioral}.
When RCTs are conducted in naturalistic real-world settings, they are called field experiments. These are often conducted in sociotechnical settings to powerful effect; one notable example from researchers at Facebook measured the degree to which different social cues influenced users' responses to Facebook ads~\cite{bakshy2012social}. 

While less common, there are some prior examples of studies that carried out sociotechnical audits by pairing algorithm audits with a randomized experiment on users. For example, one algorithm audit identifying gender and racial biases in search results subsequently performed controlled experiments that presented users with different search results~\cite{metaxa2021image}. 
Another recent example paired a fact-checking intervention field experiment with an algorithm audit that investigated the second-order effects of altered user behavior on algorithm behavior~\cite{matias2023influencing}.

\subsubsection{A/B Testing}
Designed to conduct online experiments with users of a technical system, A/B tests randomly assign users to either an `A' or `B' condition to compare outcomes. They are notable among other types of experiments because they study sociotechnical systems, can be rapidly deployed, and are low in cost (since they utilize existing users of a system rather than enrolling paid participants as in most RCTs)~\cite{kohavi2020online}.
Unlike a true field experiment, however, the goal of an A/B test is to identify the more successful of two options given some predetermined definition of success. As a result, they need not have an \textit{a priori} hypothesis, nor a treatment and control group, and their progress is often monitored in real time, an approach that is inappropriate in a formal statistical setting~\cite{johari2017ab}. 

The A/B testing method can be applied to carry out sociotechnical audits because it can be used (1)~to investigate the efficacy of a technical system by subjecting it to different user segments (akin to an algorithm audit) and (2)~to understand users by exposing them to different versions of a technical system (a user audit). 
However, the two methods have different goals and therefore suggest different implementation choices. Since accountability is a central goal of auditing, audits are conducted without internal system access. Audits led by independent third parties are the gold standard, either as a willful choice to maintain credibility or as a matter of necessity when auditors have not been granted permission to perform their audit~\cite{metaxa2021auditing}.\footnote{Some recent work has discussed methods to facilitate effective internal audits~\cite{raji_internal_auditing}, but this strategy is not always possible.} Additionally, in domains such as targeted advertising where the audit target may in fact be a conglomeration of many different systems rather than a single platform, it is simply not reasonable to assume internal access to all of the necessary systems.
Auditors, therefore, need to audit the system as part of their sociotechnical audit, unlike A/B testers who already have direct system access and understanding. These different motivations would also usually lead auditors and A/B testers to run different experiments on users, and do so with a different level of scientific rigor. 

\subsubsection{Design Interventions}
Design interventions are another form of user experimentation that tests alternative designs to investigate their impact on users, a form of user auditing. The concept of a design intervention is very broad, as ``designs'' can encapsulate any range of technical and non-technical alterations acting in virtual or physical environments, and those subject to a design intervention might be users of an online sociotechnical system, or they might be people inhabiting a physical space~\cite{crivellaro2016places}.  
Whereas descriptive methods, RCTs, field experiments, and sociotechnical audits take a scientific approach to understand users and the world, design interventions and A/B tests tend to take the more opinionated approach of a designer or engineer primarily aiming to achieve a certain outcome.
For example, recent work tested design interventions that altered users' Facebook feeds with the goal of increasing user productivity~\cite{lyngs2020hack}.
In the ads domain, recent work in collaboration with the FTC experimented with different ad labeling designs seeking to maximize user clarity~\cite{johnson2018labels}. Though design interventions typically don't include a technical algorithm audit, sociotechnical audits share a user auditing methodology that introduces careful interventions to study user outcomes.

\begin{table}[!tb]
  \centering
  \footnotesize
    \begin{tabular}{p{0.3\linewidth} | c c | c}
    \toprule
    \textbf{Method} & \textbf{Algorithm Audit} & \textbf{User Audit} & \textbf{Sociotechnical Audit}\\
    \midrule
    {Crowdsourced Audits} & {\cmark} & {(\ \cmark\ )} & {(\ \cmark\ )}\\[0.1cm]
    {Other Algorithm Audit Methods (see~\citet{sandvig2014auditing})} & {\cmark} & {\xmark} & {\xmark}\\[0.1cm]
    \midrule
    {Descriptive Methods} & {\xmark} & {(\ \cmark\ )} & {\xmark}\\[0.1cm]
    {Randomized Controlled Trials} & {\xmark} & {\cmark} & {\xmark}\\[0.1cm]
    {Field Experiments} & {\xmark} & {\cmark} & {\xmark}\\[0.1cm]
    {Design Interventions} & {\xmark} & {\cmark} & {\xmark}\\[0.1cm]
    {A/B Tests} & {(\ \cmark\ )} & {\cmark} & {(\ \cmark\ )}\\[0.1cm]
    \bottomrule
    \end{tabular}
    \caption{
        Sociotechnical Audits (STAs) are defined as the combination of both an \textit{Algorithm Audit} covering the technical aspects of a sociotechnical system and a \textit{User Audit} covering the social, user-oriented components. We situate existing methods with respect to this definition, indicating whether a method \textit{does} \cmark, \textit{does not} \xmark, or \textit{possibly} (\ \cmark\ ) includes each audit type. These methods can be adapted or combined to carry out an STA.
    }
    \label{table:intervenr_definitions}
    
\end{table}

\subsection{Sociotechnical Auditing}
To summarize, we provide our definition of a Sociotechnical Audit (or STA): a two-part audit of a sociotechnical system that consists of both an Algorithm Audit and a User Audit. We define Algorithm Audits as investigations that change inputs to an algorithmic system (e.g., testing for a range of users or behaviors) and observe the system's outputs to infer properties of the system. Meanwhile, we define User Audits as investigations that change inputs to the user (e.g., different system outputs) and observe their effects to draw conclusions about users. We summarize the methods covered in this section in Table~\ref{table:intervenr_definitions}.

\section{Intervenr: A System for Sociotechnical Auditing}
\label{section:system}
Having introduced the concept of sociotechnical auditing, our second contribution is Intervenr, a web browser-based system for running sociotechnical audits---including but not limited to our targeted ads audit.

\subsection{Our Sociotechnical Audit Approach}
\begin{figure}[!tb]
  \includegraphics[width=\textwidth]{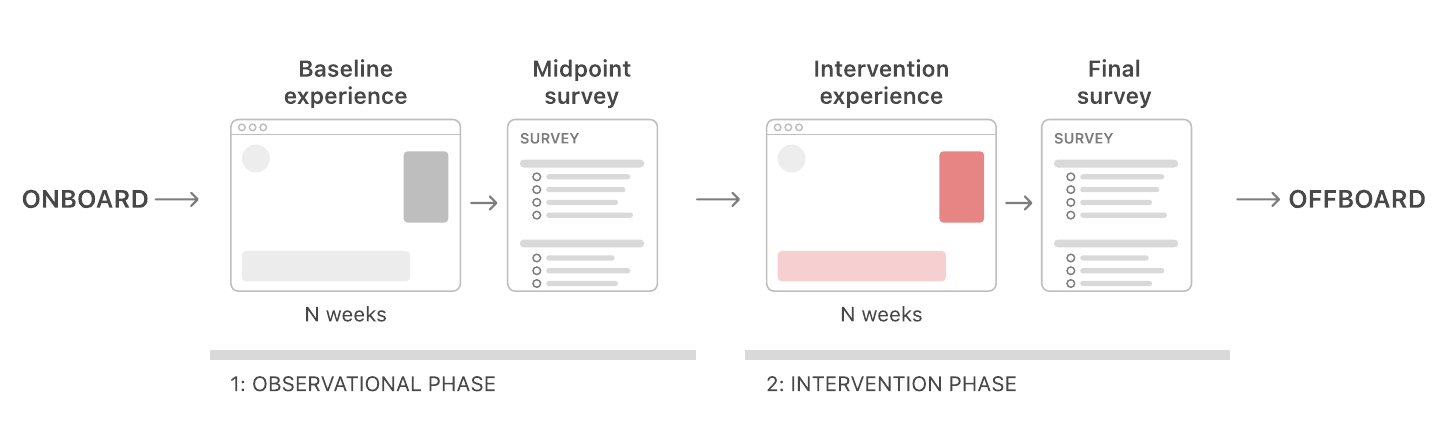}
  \caption{
    Our sociotechnical audit study design.
    Participants first onboard to the study and then enter the \textit{Observational Phase}. During this phase, the system captures their baseline experience and conducts a midpoint survey on that experience. Then, participants enter the \textit{Intervention Phase}, in which the system enacts the intervention and conducts a final survey, after which participants are offboarded and compensated.
  }
  \label{fig:intervenr_study_design}
\end{figure}

Sociotechnical audits investigate both the technical and human components of a sociotechnical system as they interact. We instantiate this method with a two-phase study design: an observational phase to audit the algorithmic component, followed by an intervention phase to audit the user component of the sociotechnical system~(Figure~\ref{fig:intervenr_study_design}). 
During the observational phase, we collect observational data (including algorithmic content and user behavioral data) from participants. At the conclusion of this phase, we conduct a midpoint survey to gather baseline measurements. The initial observational phase audits the \textit{technical} side of the system by analyzing outcomes from a range of users for the status quo algorithm.
Then, during the intervention phase, we enact \textit{in situ} interventions that modify the algorithm's behavior and collect similar data, concluding with a final survey that gathers post-intervention measurements. The intervention phase audits the \textit{human} side of the system by comparing baseline and post-intervention data to understand how changes in the algorithm lead to changes in user outcomes.

When designing a sociotechnical audit with our study design, the auditor can freely alter:
\begin{itemize}
    \item Data collection: The kind of data automatically collected through the system, including digital media and metadata as well user actions, such as clicks and views.
    \item Survey design: The questions included in the midpoint and final surveys, which can draw upon data collected from participants in either study phase.
    \item Intervention design: The logic to alter, replace, or otherwise intervene upon the media that users encounter.
    \item Participant pool: The set of users to study, which can be curated with custom sign-up surveys that sample according to relevant factors of participant representativity.
    \item Study timing: The duration of study phases and timing of surveys and interventions.
\end{itemize}

\begin{figure}[!tb]
  \includegraphics[width=\textwidth]{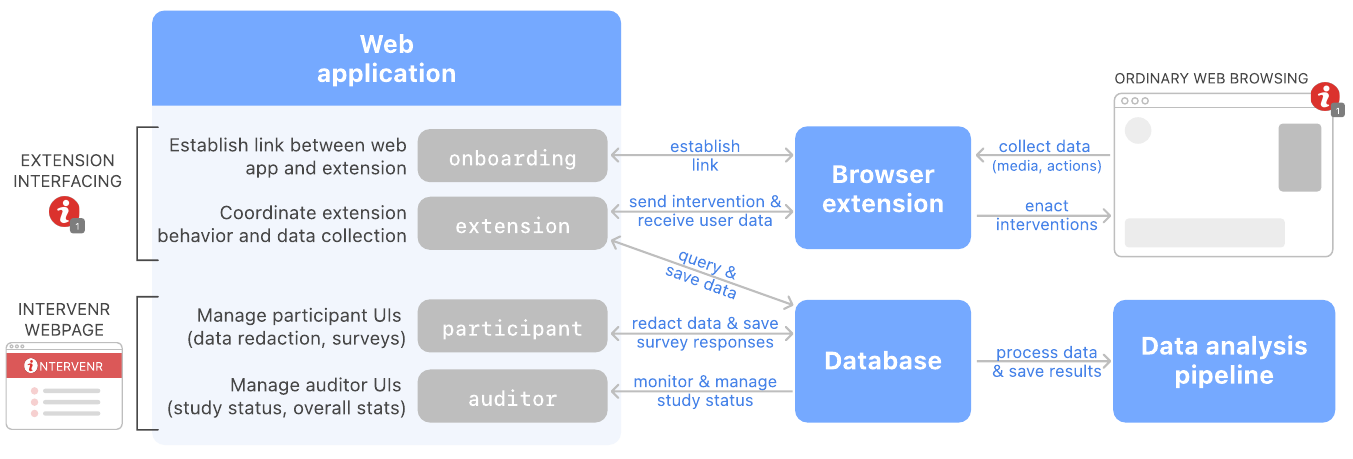}
  \caption{
    The Intervenr System for sociotechnical audits. 
    \textit{Web application}---Implements interfaces to manage auditors and study participants. Coordinates with the browser extension to collect data and enact interventions.
    \textit{Browser extension}---Acts during participants' ordinary web browsing to collect media and user actions as well as perform in situ interventions on webpages.
    \textit{Database and data analysis pipeline}---Stores collected media, survey responses, and post-processing results. Performs offline processing of the data collected in the audit.
  }
  \label{fig:intervenr_system}
\end{figure}

To conduct our STA study design, we built a system called \textit{Intervenr} that consists of a browser extension and web application as well as an accompanying data analysis pipeline~(Figure~\ref{fig:intervenr_system}). This three-part system allows us to effectively collect user data and enact interventions while providing support for study participants and auditors.

\subsection{Design Goals}
To enact sociotechnical audits requires a flexible approach that studies both the technical and human components of a sociotechnical system.
Central to this auditing approach, our system needed to not only support passive observation of users' experiences with algorithms, but also provide \textit{active intervention} capabilities to modify those experiences and audit the users.
In our audit case study, we focused on in-browser targeted advertising, not restricted to any specific website or ad source, so our system needed to be sufficiently robust to collect many \textit{different types of data} originating from the \textit{full range of sites} that users encounter in their daily browsing.
Finally, because our audit was designed to capture a diverse group of users' experiences in great detail, it needed to support the \textit{usability and privacy needs} of a broad range of users.

\subsection{The Intervenr System}
Motivated by our design goals, we chose the web browser to instantiate our tool for several key reasons.
First, web browsers capture a large segment of a user's digital media exposure within a single application. 
Since STAs rely on user participation, we sought to minimize the onboarding complexity and invasiveness of our system. Intervening upon the behavior of a single user-facing application allows us to provide a streamlined participant onboarding process. 
Web browsers also provide a comprehensive suite of instrumentation to collect user behavior and modify existing sites in situ, both of which were critical to executing a sociotechnical audit. 
Finally, one goal of our system is to provide a system that auditors can extend to run a variety of STAs; a web development stack is accessible to a broader set of auditors, granting them the flexibility to customize their interventions. A major limitation of this approach is our inability to collect mobile data, which we elaborate upon in the Limitations section (see Section \ref{sec:discussion}).

\begin{figure}[!tb]
  \includegraphics[width=\textwidth]{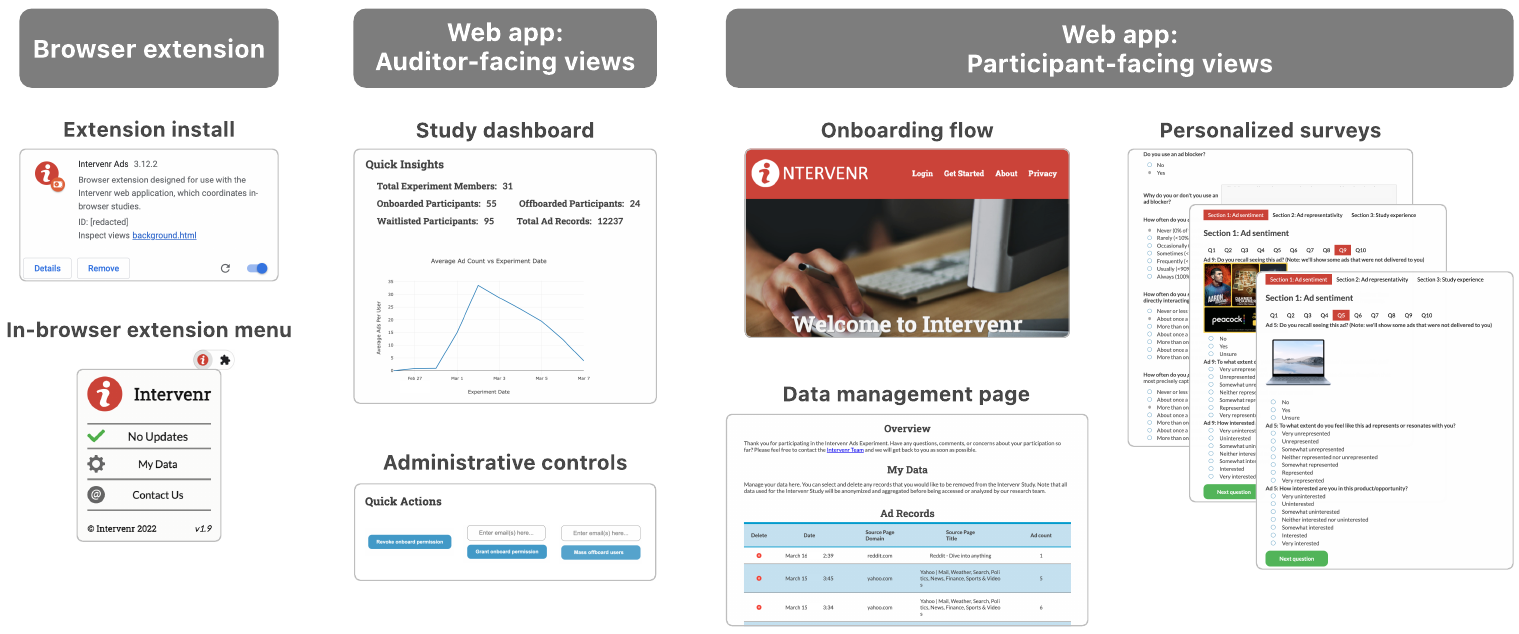}
  \caption{
    The Intervenr browser extension and web application interfaces include auditor-facing views (study dashboard and admin controls) and participant-facing views (onboarding, data management, and surveys).
  }
  \label{fig:intervenr_system_screens}
\end{figure}

Next we provide an overview of the key components of the Intervenr system (Figure~\ref{fig:intervenr_system_screens}).

\subsubsection*{Browser extension}
With our focus on algorithms that users encounter via web browsing, we turn to a custom browser extension to support fine-grained observation and intervention across the web. Extensions are installed at the browser level and remain active across all of a user's web browsing behavior.
During the observational phase, Intervenr's browser extension allows researchers to gather granular information about the content that users encounter as well as their behavior (for example, in terms of views and clicks). Then, during the intervention phase, the extension allows researchers to enact fine-grained, in situ interventions on the webpages that users visit.

\subsubsection*{Web application: Central experiment server and database}
While a browser extension is ideal for carrying out the mechanics of user data collection, STAs must also enact interventions that are centrally designed and coordinated by an auditor. Toward this end, our system includes a central experiment server that manages experiment details and deploys the logic to ensure that the user-facing browser extension collects the right data and enacts the right interventions. This server is also responsible for storing the data collected during the audit to facilitate analysis.

\subsubsection*{Web application: Participant-facing interface}
Then, since these audits involve longitudinal data collection and active participant involvement via surveys, our system includes a dedicated participant-facing interface. Participants can use this interface to complete the full study lifecycle: signing up, providing consent, onboarding, participating in the study, and offboarding. During their study participation, this interface provides key information and instructions on study involvement, and at specified points, hosts personalized surveys assigned to the participant. This interface also provides comprehensive data management controls where users can view all of the data that our system has collected. At any point during the study, they can redact any data items that they would not like to share. Our system deletes the data from the database accordingly and only logs metadata about the number of records that a user has redacted.

\subsubsection*{Web application: Auditor-facing interface}
Sociotechnical audits are meant to support both a large number of participants and a longitudinal, continuous study deployment. Thus, our system provides an auditor-facing interface to assist auditors in facilitating and monitoring their large-scale studies over time. This interface provides an overview of the data collected by the system as well as administrative controls to perform key actions like granting onboarding permission to users, launching participant surveys, monitoring survey completion, and offboarding users.

\subsubsection*{Data analysis pipeline}
Finally, because the goal of an STA is to ultimately evaluate statistical claims about system behavior, our system is designed to support large-scale analysis of participant data. Our data analysis pipeline consists of scripts that auditors can author to run at specified intervals to augment their analysis. For example, nightly scripts might download persistent versions of images from their URLs, which might go stale. Weekly scripts might run on images to perform additional automated processing such as optical character recognition (OCR), object detection, or person recognition.

\subsection{Intervenr Participant Experience}
To provide a clearer picture of these system components in practice, we walk through a study participant’s experience of Intervenr.

\textit{Signing up}. Beginning on the Intervenr homepage, participants are added to a waitlist after providing demographic and contact information, and agreeing to the study's consent. 

\textit{Onboarding}. Participants selected by the researcher are guided through the process of setting up their account on the website, installing the browser extension, and linking the two. 

\textit{Managing their data}. After onboarding, data collection begins automatically while participants browse the web as usual. They can visit the website to view all data collected from them and redact any that they choose. Participants can also exclude data by browsing in an incognito window. 

\textit{Completing surveys}. Participants periodically receive emails to complete surveys on the Intervenr site via personalized, authenticated links. These auditor-designed surveys can contain custom questions for each participant, including questions that draw on their own personal data. 

\textit{Offboarding}. To offboard at any time, participants can simply remove the extension and log out from the website, terminating data collection. Compensation is distributed through online gift cards sent by email, and their data is eventually deleted at the end of the audit project.

\subsection{Intervenr Auditor Experience}
We briefly walk through an auditor’s experience working with Intervenr to design and execute a sociotechnical audit.

\textit{Designing the study}. First, the auditor must design their study (including deciding what browser content to collect, how long to run the study, what interventions and surveys to deploy, etc.), and adapt the current Intervenr system accordingly.

\textit{Launching the study}. When they are ready, auditors can launch their Intervenr study and begin recruiting participants as they see fit by directing would-be participants to the website.

\textit{Running the study}. Once the desired participant pool has been achieved, the auditor can begin the study, monitoring its progress using the admin dashboard, releasing surveys, filtering participants, and deploying interventions with the help of the platform. Only compensation at the end of the study is not directly handled by the system and must be coordinated separately.

\textit{Analyzing study results}. After the study, auditors can conduct their data analysis using computational notebooks or custom scripts to query the study database where browsing data and survey results are stored.

\subsection{Implementation Details}
We implement the browser extension as a Google Chrome extension using Manifest Version 2 and the web application as a Django web app deployed on Heroku. The data analysis pipeline is set up on an Amazon EC2 server with several custom Python scripts configured to run at periodic time intervals. Please see Appendix~\ref{appendix:implem-details} for details on our Intervenr infrastructure implementation.

\section{Case Study: A Sociotechnical Audit of Targeted Advertising}
\label{section:study}
We use Intervenr to perform a sociotechnical audit on a complex algorithmic ecosystem that heavily impacts users: targeted online advertising. This is an especially relevant area of study because such ads are explicitly designed to shape user beliefs. Contrary to other forms of media like search results, social media, or news, online advertising grants very little user control: the ads that users see are almost entirely curated by algorithmic systems rather than selected by users. Further, targeted online advertising is fundamental to the profitability of many companies, and insights leading to changes or regulation would have potential for large impact~\cite{cnbc2019digitalAdRevenue}. Given this combination of strong profit motives, near-total algorithmic curation, and highly persuasive media shaping user beliefs and behaviors, targeted advertising is a high-stakes domain for users, technology companies, and society at large.

As we outlined earlier, online advertising is an especially challenging form of digital media to study using existing evaluation strategies. Since it is heavily personalized and involves repeated exposures across many websites and over time, controlled lab studies hold little ecological validity. Audits that passively collect real user data come up short without the ability to test interventions. Online ads are also designed for passive, unconscious consumption. Since users often do not actively attend to ads~\cite{resnick2014adBlindness}, they may struggle to recall the ads they've been exposed to in surveys, let alone the broader impact that ads have on their behaviors and beliefs.

Our Intervenr system directly addresses these challenges, as the system is explicitly designed to capture and intervene on the personalized media that participants encounter in-the-wild, across sites, and over longer stretches of time. Pairing our data collection with participant behavior traces and personalized surveys, we are able to identify and study advertisements that users do not recall, and we can ask participants in-depth questions about specific ads we know were delivered to them.

\subsection{Our Ad Targeting Intervention}
The goal of our intervention is to investigate the efficacy of targeted online advertising according to a number of metrics. We design an \textit{ablation-style intervention} that compares the status quo of normal targeted ads to a version with broken or disrupted ad targeting. Just as ablation studies systematically remove components of a system to understand how they contribute to overall performance~\cite{meyes2019ablation}, our ablation intervention disrupts normal ad targeting to help us understand how it contributes to the functioning of ad systems.
Without internal access to ad targeting algorithms, our audit instead relies on our ability to change how ads are delivered in our participants' browsers. With this approach, we can break ad targeting at the user-level by swapping ads between users.

\subsubsection{Intervention Design Goals}
While there are many ways auditors might break ad targeting algorithms, we design our intervention to simultaneously preserve two important qualities of targeted ads.
First, we aim to preserve the \textit{ecological validity of ads}.
By swapping all of the ads between users, we maintain the validity of the ad sets (compared to an intervention that serves all users an entirely random set of ads). 
Each of the ad sets are realistic because they were in fact delivered to another real user, allowing us to isolate the role of \textit{personalized} targeting. 

\begin{figure}[!tb]
  \includegraphics[width=0.6\textwidth]{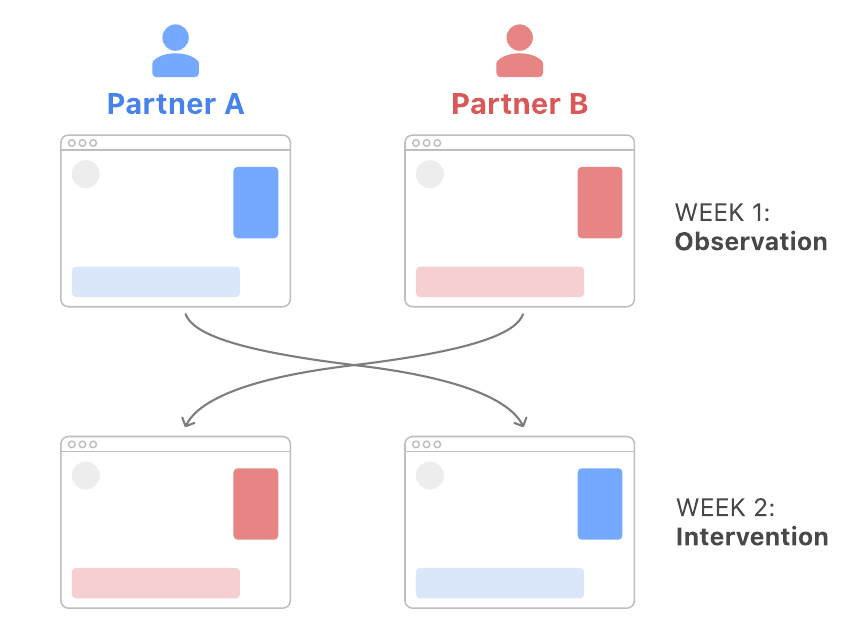}
  \caption{
    Ad-swapping intervention. We first randomly assign participants a partner. In the observational phase, participants receive online ads as usual. In the intervention phase, we swap all ads between the partners so that Partner A will only see ads targeted to Partner B, and vice versa.
  }
  \label{fig:intervenr_study_ads}
\end{figure}

Second, we seek \textit{sufficient variation in paired user similarity}. Part of the insight behind our ad-swapping intervention is that there are varied impacts on different pairs of users depending how similar they are to each other. 
For two very similar users, swapping ads minimally disrupts ad targeting, but for users who are very different from one another, swapping brings about a significant disruption. 
To study the impact of disrupted targeting along a continuum, we use \textit{random pairing}. With a sufficiently-sized sample, this produces natural variation in user similarity, even with respect to factors we may not have been able to account for upfront; this is the same role that randomization plays in randomized controlled trials. 

\subsubsection{Final Intervention Setup}
Our intervention acts on a \textit{user} level to swap ads between randomly-assigned pairs of participants (Figure~\ref{fig:intervenr_study_ads}). At the start of the study, users are paired; we term these ``swap partners.''
During the observational phase, users see their normal targeted ads without any interference, and Intervenr collects these ads. 
Then, during the intervention phase, users exclusively see their assigned swap partner's ads. Each time an ad renders on their web browser, Intervenr replaces it with an ad sampled from their swap partner's collected ad set.

Throughout this paper, we use the term ``targeting'' to refer to the ads that are delivered to a particular user, which include ads that are directly targeted to the user (using targeting data such as demographics and interests) as well as ads that were not targeted based on such data. We do not have access to the targeting data that advertisers hold and thus cannot assess the quality of this data and its impact on ad delivery. However, our broader definition of targeting allows us to ask a larger question with our intervention: considering \textit{all} of the ads that a user in fact sees in their daily life, to what extent does it matter that they're receiving those ads and not another user's ads?

\subsection{Research Questions}
Our case study aims to answer several key questions related to the efficacy of targeted advertising, in terms of outcome metrics that align with user experiences as well as metrics aligned with advertiser goals. 
Marketing research has traditionally focused on advertiser-oriented metrics that emphasize the revenue accrued by advertisements. Given the rarity of conversions (i.e., purchases) among ad placements, advertisers primarily optimize for user engagement (views, clicks) as a proxy for user interest. However, even user engagement events like ad clicks are very sparse; prior work has estimated click rates as low as 0.05\% to 0.1\% for display ads online~\cite{Aribarg2020NativeAdvertising, tucker2014Advertising}. 
This exceedingly low engagement rate and advertiser-centric focus motivated our choice of alternative, user-centric metrics that could directly capture a user's perceptions of ads.

A user's own stated interest in an ad is of course a primary metric we must consider. In addition, scholars have long reminded us that advertisements do much more than display products---ads are often tailored based on the expected identity characteristics of the receiver so as to shift their beliefs and self-perceptions in directions favorable to the advertiser~\cite{goffman1979gender}, and often in ways that may perpetuate harmful bias and discrimination~\cite{mcilwain2023algorithmicPL, metaxa2023biasPL}. Recent research has presented evidence that ads are often delivered disproportionately to users with a similar identity (e.g., race, gender, age) to those pictured, likely to increase users' feelings of representativity~\cite{kaplan2022impliedIdentityAds}. Thus, users' perceptions of the representativity of ads serve as an important signal that may point to the broader belief-level impact of ads.

Thus, in our study, we focus more heavily on our user-aligned metrics: (1)~\textit{interest}: the extent to which a user expresses interest in an ad and (2)~\textit{representativity}: the extent to which a user feels represented by an ad. 
For consistency with prior work, we secondarily investigate several advertiser-aligned metrics: (1)~\textit{recognition}: the proportion of ads that users correctly recognize, (2)~\textit{views}: whether ads are seen by users, as measured by ads that enter the user-visible viewport, and (3)~\textit{clicks}: the proportion of ads that users click.
These inform our core research questions:

\begin{enumerate}[leftmargin=1.25cm, rightmargin=1cm, label=\textbf{RQ\arabic*:}]
    \itemsep=10pt
    \item How well does the current ad targeting system work (an algorithm audit)?
    \begin{enumerate}[leftmargin=1cm]
        \itemsep=0pt
        \item [\textit{RQ1-U}:] How well does current ad targeting work for \textit{users}, in terms of their ad interest and feeling of representation?
        \item [\textit{RQ1-A}:] How well does current ad targeting work for \textit{advertisers}, in terms of ad recognition, views, and clicks?
        \item [\textit{RQ1-D}:] How do the above results vary by user race, gender, and other \textit{demographics}?
    \end{enumerate}

    \item How do users respond to personalized ad targeting relative to an alternative ad delivery method that disrupts targeting (a user audit)?
    \begin{enumerate}[leftmargin=1cm]
        \itemsep=0pt
        \item [\textit{RQ2-U}:] How much relative impact does ad targeting have on \textit{users}, in terms of their ad interest and feeling of representation?
        \item [\textit{RQ2-A}:] How much relative impact does ad targeting have on \textit{advertisers}, in terms of ad recognition, views, and clicks?
        \item [\textit{RQ2-D}:] How do the above results vary by user race, gender, and other \textit{demographics}?
    \end{enumerate}
\end{enumerate}

\subsection{Study Design}
Next, we walk through the details of our study design, including the recruitment procedure, study experience, and compensation.

\subsubsection{Participant Recruitment} 
Since we are interested in the differing impact of online ads on users of different demographics, we aimed to recruit a diverse set of participants, especially with respect to race and gender. 
We conducted our recruitment through Prolific, a platform for sourcing research participants. There, we hosted a short 1-minute screener survey relating to eligibility requirements.\footnote{To be eligible, users needed to live in the U.S., be 18 years or older, regularly use a laptop or desktop computer, and use Google Chrome as their main web browser.} Eligible participants were directed to a page notifying them of the study. We made clear that payment for the screener was not related to the Intervenr study, and that the Intervenr study was an externally-hosted study not affiliated with Prolific. All participants were compensated \$0.25 (a \$15/hour rate) for completing the screener, regardless of main study eligibility.

Out of the 5,600 people who completed our screener, 1,310 (23.4\%) signed up for the Intervenr study. Based on a power analysis on earlier pilot data, we targeted to include at least 200 participants in our study, and allocated 600 spots for onboarding in anticipation of study attrition. To balance participants by race and gender, we selected all participants who did not identify as white (N=247), and then randomly sampled from the remaining pool of white participants while balancing the number of men with the number of marginalized-gender participants (those identifying as women and/or non-binary). Of the 600 selected participants, a total of 402 people (67.0\%) completed the full onboarding process. 
During onboarding, participants completed a background information survey including demographic information such as their age, race, gender, location, education level, and household income. We report on the demographics of our participant pool in Section~\ref{section:results-overview}.

\subsubsection{Study Phases} 
During the observational phase, the user experience was unaltered while our extension recorded all ads that appeared on the web pages participants loaded in their browsers, along with clicks and views on those ads.

During the intervention phase, users received ads that were originally delivered to their random swap partner. Our extension recorded all original ads that would have normally been delivered to the user, as well as the swap-partner ads that replaced them. Our system also recorded clicks and views for the swap ads shown.

\subsubsection{Compensation and Participant Management} 
Participants were compensated \$10 for completing each milestone of the study that required their active effort: onboarding, the midpoint survey, and the final survey. Since each of these action items required about 10-15 minutes to complete, this amounted to a rate of at least \$40/hour. Participants received compensation in the form of an Amazon gift card.  

We communicated with participants over email throughout the course of the study to notify them of onboarding permission, survey deployments, and offboarding. We sent an additional email on the fourth day of each study phase notifying participants with no ads collected to check their system setup, and assisted users over email to debug any setup issues.

\subsection{Survey Design} 
\label{section:survey-design}
Both the midpoint and final surveys consisted of holistic ad questions to capture broader ad impressions, per-ad questions to capture more granular feedback, and study experience questions. 

\subsubsection{Holistic ad questions}
We began each survey by showing participants an image cloud with a random sample of up to 40 ads they had seen in the week prior, and asked them to indicate their recognition (the approximate proportion of ads that they recognized from this visual),\footnote{We asked ``What proportion of these ads do you recall seeing?'' and participants selected from among seven percentage options ranging from 0 to 100\%.} interest (whether they were interested in the ads), and feeling of representation (whether they felt the ads represented them). The latter two items were rated on a 7-point Likert scale.

\subsubsection{Per-ad questions} 
In the per-ad question section, we presented participants one ad image at a time and asked questions about recognition (whether they remembered seeing it),\footnote{We asked ``Do you recall seeing this ad?'' and participants could select a \textit{No}, \textit{Yes}, or \textit{Unsure} option.} interest (whether they were interested in it), and feeling of representation (whether they felt it represented them). The latter two items were again rated on a 7-point Likert scale.

To select ads for the per-ad survey section, we sampled to cover three factors:
\begin{itemize}
    \item \textbf{Targeted-user}. Whether the ad was originally targeted to the participant (``self'') or was originally targeted to their partner. This was central to our intervention; we sought to capture user responses to both types of ads (those targeted to themselves and to their partner) both before and after the intervention. Valid values: ``self'' or ``partner.''
    \item \textbf{Seen-status}. Whether an ad was seen by the participant or not. This was measured by tracking which ads entered the user's visible view on screen, in order to account for the impact of actually seeing an ad. In other words, ``unseen'' ads are those that were loaded on the user's webpage, but that the user did not scroll down far enough to see. Valid values: ``seen'' or ``unseen.''
    \item \textbf{Has-people}. Whether the ad contained at least one person. This was determined using an automated person detection model. We included this factor because prior work suggests people may respond differently to imagery of other people~\cite{strahan2006comparing}. Valid values: ``people'' or ``noPeople.''
\end{itemize}

For each study phase, there were six possible combinations of values for the three factors (Table~\ref{table:per_ad_categories}), as there was a forced mapping between seen-status and targeted-user: during the observational phase, only self ads can be seen, and all partner ads are unseen; during the intervention phase, only partner ads can be seen, and all self ads are unseen. We sampled up to four ads from each category for each participant, such that participants received up to 24 per-ad questions. 

For the observational phase, this resulted in the following six categories: (1)~seen-self-people, (2)~seen-self-noPeople, (3)~unseen-self-people, (4)~unseen-self-noPeople, (5)~unseen-partner-people, and (6)~unseen-partner-noPeople.
For the intervention phase, this provided the following six categories: (1)~seen-partner-people, (2)~seen-partner-noPeople, (3)~unseen-self-people, (4)~unseen-self-noPeople, (5)~unseen-partner-people, and (6)~unseen-partner-noPeople.

\begin{table*}[!tb]
  \centering
  \scriptsize
    \begin{tabular}{r l l l}
    \toprule
    \textbf{Study Phase} & \textbf{Seen-status} & \textbf{Targeted-user} & \textbf{Has-people}\\
    \midrule
    \multirow{6}{1.5cm}{\raggedleft\textit{Observational}} & {seen} & {self} & {people}\\
    & {seen} & {self} & {noPeople}\\
    & {unseen} & {self} & {people}\\
    & {unseen} & {self} & {noPeople}\\
    & {unseen} & {partner} & {people}\\
    & {unseen} & {partner} & {noPeople}\\
    \midrule
    \multirow{6}{1.5cm}{\raggedleft\textit{Intervention}} & {seen} & {partner} & {people}\\
    & {seen} & {partner} & {noPeople}\\
    & {unseen} & {self} & {people}\\
    & {unseen} & {self} & {noPeople}\\
    & {unseen} & {partner} & {people}\\
    & {unseen} & {partner} & {noPeople}\\
    \bottomrule
    \end{tabular}
    \caption{
        A summary of the categories of ads that were sampled for the per-ad survey questions in each study phase. There were six categories for each phase (\textit{seen} ads in each phase could only belong to one of the targeted-user options: \textit{self} in the observational phase, \textit{partner} in the intervention phase).
    }
    \label{table:per_ad_categories}
\end{table*}

\subsubsection{Study experience questions}
The last section of each survey contained a short set of questions related to participants' overall experience with the study so far. Participants were asked to rate their experience and the likelihood that they would recommend the study to a friend. They were also asked compliance-related questions to report the frequency with which they had disabled our extension or used incognito mode to circumvent data collection. We also provided a free-text field for any additional comments.

\subsection{Ad-Specific Infrastructure}
To support our ad targeting case study, we implemented several modifications to the core Intervenr infrastructure. First, we set up an ads-specific browser extension that integrated an open-source ad blocker extension, AdNauseam,\footnote{AdNauseam is an open-source browser extension~\cite{nissenbaum2009trackmenot} that in turn builds on uBlock Origin, a content-blocking extension that blocks ads, trackers, and malware sites.} and added functionality to dynamically remove and insert ads. We additionally added functionality to track user ad views and clicks. Then, we extended the web application to execute our ad-swapping intervention, participant surveys, and auditor dashboard. Finally, we extended our data analysis pipeline to perform ad link resolving, ad image downloading, and automated person detection for all ad images. See Appendix~\ref{appendix:ad-infra-details} for implementation details.

\section{Case Study Results}
\label{section:results}
Our two-week sociotechnical audit on targeted advertising ($N=244$ users completing the full study) found that ad targeting appears to have a substantial impact on users' ad recognition and perceptions of ad interest and representativity. When our ad-swapping intervention systematically broke targeting, there were significant decreases across ad metrics, and marginalized users appeared to be more negatively impacted by this disruption. 
However, after just one week of exposure, we observed an increase in user affinity for their random swap partner's ads, and a slight decrease in interest towards their own ads.
These findings suggest that a substantial part of the efficacy of ad targeting may be driven by providing repeated exposures that foster user familiarity, rather than by providing inherent value to users through personalized targeting.

Following the structure of our sociotechnical audit definition, we first report the results of our algorithm-focused audit, followed by the user-focused findings. We conclude with basic information about our participants, the collected advertisement data, and steps we took to verify the quality of our analyses.

\subsection{Auditing the Algorithm (RQ1)}
First, we first report the results of our algorithm-focused audit on the performance of current personalized ad targeting.

\subsubsection{Personalized ads perform moderately (RQ1-U, RQ1-A)}
Our results show that personalized ads performed only moderately for users and advertisers.
Users' baseline levels of ad interest and representativity were moderate and relatively close to neutral on our 7-point Likert scale.\footnote{For any given attribute X, our 7-point Likert scale ranged from ``1: Very un-X'' through ``7: Very X'' (i.e., ``1: Very uninterested'' through ``7: Very interested''), with a score of 4 corresponding to a neutral response.} For the holistic ad survey question that displayed a cloud of ad images actually seen by the participant, the mean ad interest was $3.89$ ($SD=1.87$); the mean ad representativity was $4.10$ ($SD=1.67$)~(Figure~\ref{fig:interv_holistic_interestRep}). The per-ad survey questions, which gathered responses to individual ads, similarly reflected moderate ad interest ($M=3.72, SD=1.38$), and ad representativity ($M=3.90, SD=1.33$).

We found that ad recognition was generally quite low at around $30-50\%$. For the holistic ad survey question, which showed an image cloud of ads they had in fact seen, users self-estimated their recognition to be between 30\% and 50\% ($M=3.27, SD=1.64$ on our Likert scale).
Similarly, for the per-ad survey questions, participants correctly recognized $40.9\%$ of ads ($SD=24.5\%$) that they had actually seen (Figure~\ref{fig:perAd_recall}).
Out of all ads delivered in the observational phase, on average 27.2\% were viewed ($SD=14.9\%$), and a very low percentage of viewed ads were clicked ($M=0.057\%, SD=0.22\%$).\footnote{This figure is consistent with prior work measuring 0.05\% to 0.1\% average click rates for display ads~\cite{Aribarg2020NativeAdvertising, tucker2014Advertising}.}

\subsubsection{Ad interest and representativity differs by participant race (RQ1-D)}
\begin{figure}[!tb]
  \includegraphics[width=0.8\textwidth]{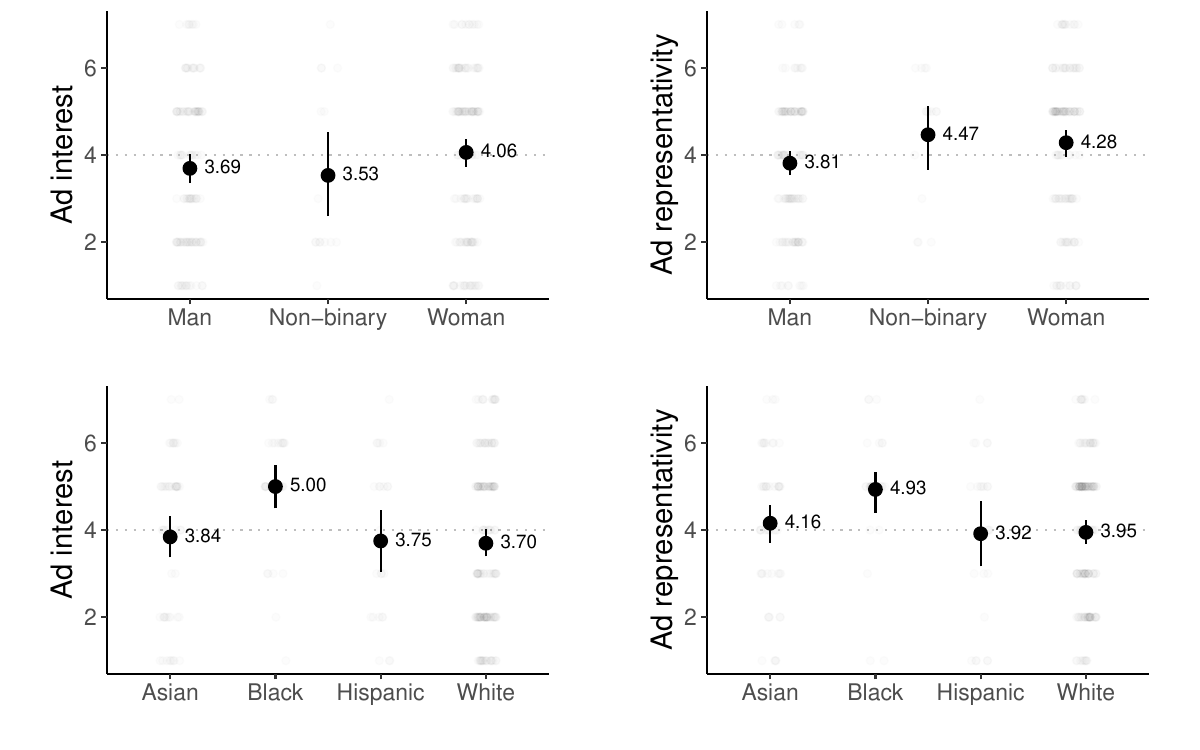}
  \caption{
    Baseline ad interest (left) and ad representativity (right) from the observational phase, broken down by participant gender (top) and race (bottom).
    Ad interest and representativity were rated on 7-pt Likert scales.
  }
  \label{fig:obs_holistic_genderRace_interestRep}
\end{figure}

We found significant differences in ad interest and representativity based on participant race---in particular, that Black participants responded more positively to their targeted ads.
To analyze all of our survey metrics---ad interest, representativity, and recognition---on the holistic ad question with respect to participant demographics, we used a linear regression model with fixed effects of user age, education, income, geographical region, and the interaction between race and gender.\footnote{\texttt{metric\_val \textasciitilde\ 1 + age + education + income + region + (race*gender)}} 
We did not observe any significant effects among these variables on recognition, but we did observe a significant effect of race on ad interest ($F(3, 221)=3.73, p < 0.05$), as well as representativity ($F(3, 221)=2.97, p < 0.05$). In particular, as visualized in Figure~\ref{fig:obs_holistic_genderRace_interestRep}, Black participants reported higher responses than others on these metrics. 

\subsubsection{Ad performance differs depending on whether ads were seen, targeted to the user, or contained people (RQ1-U, RQ1-A)}
\label{sec:res:seen-status}
For the per-ad survey questions, we observed significant effects based on all three labeled attributes of the ads (seen-status, targeted-user, and has-people) for all three survey metrics (interest, representativity, and recognition).
We used a linear mixed-effects model with random effects of participant ID and fixed effects of seen-status, targeted-user, has-people, age, education, income, geographical region, and the interaction between race and gender.\footnote{\texttt{metric\_val \textasciitilde\ 1 + (1 | participant\_id) + seen + targeted\_user + has\_people + age + education + income + region + (race*gender)}}
We used this alternative model formulation compared to the previous model to control for individual differences between participants, since we had repeated measures for each participant.
Seen-status ($F(1, 5247)=26.28, p < 0.001$), targeted-user ($F(1, 5261)=35.97, p < 0.001$), and has-people ($F(1, 5260)=97.31, p < 0.001$) all had significant effects on users' reported interest in ads. 
The same was true for ad representativity: seen-status ($F(1, 5251)=23.70, p < 0.001$), targeted-user ($F(1, 5265)=45.90, p < 0.001$), and has-people ($F(1, 5264)=97.45, p < 0.001$) were all significant.
This was also the case for ad recognition: seen-status ($F(1, 6262)=31.27, p < 0.001$), targeted-user ($F(1, 5283)=70.05, p < 0.001$), and has-people ($F(1, 5282)=68.25, p < 0.001$) had significant effects.

Diving further into these three factors post-hoc, we observed higher ad interest, representativity and recognition for (1)~ads that \textit{were seen} over ads that were not seen by the user (Figure~\ref{fig:obs_interv_perAd_seenStatus}), (2)~ads that were \textit{originally targeted to the user} over ads that were targeted to their random swap partner (Figure~\ref{fig:obs_interv_perAd_targetedStatus}), and (3)~ads that \textit{do not contain people} over ads that do contain people (Figure~\ref{fig:obs_interv_perAd_pplStatus}).

\subsection{Auditing the Users (RQ2)}
Next, we report on the user-focused findings enabled by our sociotechnical auditing method, which examine how users respond to an alternative ad delivery method that disrupts status quo targeting.

\subsubsection{Personalized targeted ads outperform swapped ads (RQ2-U, RQ2-A)}
\begin{figure}[!tb]
  \includegraphics[width=0.7\textwidth]{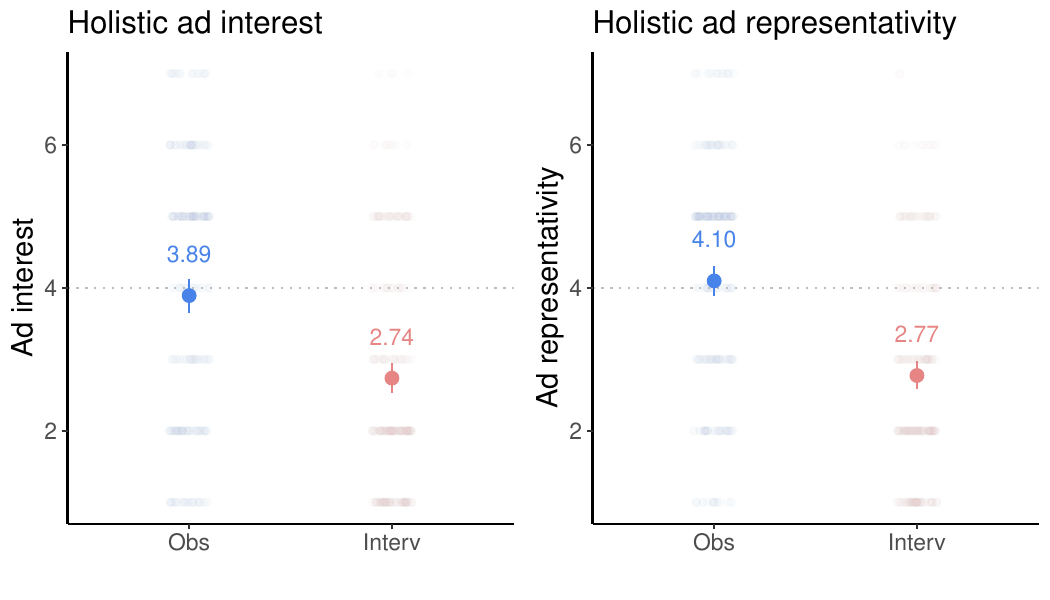}
  \caption{
    Holistic ad interest and representativity across study phases (7-pt Likert scale). Interest and representativity were near neutral in the observational phase, but dropped substantially in the intervention phase.
  }
  \label{fig:interv_holistic_interestRep}
\end{figure}

Our ad swapping intervention resulted in substantial decreases in both ad interest and representativity (Figure~\ref{fig:interv_holistic_interestRep}). 
For the holistic ad survey question, we conducted one-sided paired $t$-tests comparing participants' own personalized ads during the observational phase to non-personalized swap ads seen in the intervention phase. We found a significant metric drop for ad interest, from an average rating of $3.89$ to $2.74$ ($t(243) = 8.73, p < 0.001$, Cohen's $d = 0.56$) and ad representativity, from an average of $4.10$ to $2.77$ ($t(243) = 9.79, p < 0.001$, Cohen's $d = 0.63$). 

We observed similar trends in the per-ad questions. One-sided paired $t$-tests showed significantly lower interest in swap partners' ads ($M=2.93, SD=1.36$) compared to one's own ($M=3.72, SD=1.38$); $t(239) = 7.96, p < 0.001$, Cohen's $d = 0.41$. We saw a comparable drop in ad representativity between one's own ads ($M=3.90, SD=1.33$) versus swapped ads ($M=3.14, SD=1.34$); $t(239) = 7.38, p < 0.001$, Cohen's $d = 0.40$.

We found that self-estimated ad recognition on the holistic survey question dropped significantly from the observational phase ($M=3.27, SD=1.64$) to the intervention phase ($M=2.46, SD=1.50$); $t(243) = 6.43, p < 0.001$, Cohen's $d = 0.41$.
However, on per-ad survey questions, correct recognition remained relatively consistent ($M=43.0\%, SD=25.7\%$) compared to the observational phase ($M=40.9\%, SD=24.5\%$) (Figure~\ref{fig:perAd_recall}), and a two-sided $t$-test found no significant difference between the correct recognition rates in the two phases ($t(252.6)=-0.70, p = 0.49$).
One possible explanation for this discrepancy is that when presented with a high-level view of ads in the holistic question, participants may have noticed that the ads were not their typical ads (i.e., that these ads were their swap partner's ads) and may have reported a lower recognition rate for these unfamiliar-looking ads, and thus this self-report was not borne out by their responses to specific ads. 

\subsubsection{Users acclimate to ads after exposure (RQ2-U, RQ2-A)}
\label{section:res-interv-perAd}
While our ablation intervention resulted in decreased ad performance, we found that swap partners' ads performed better with users after the end of the 1-week intervention period compared to before the intervention period began.
To analyze the per-ad survey question across study phases, we first conduct omnibus tests, using a linear mixed-effects model with a fixed effect of study phase (either observational or intervention) in addition to other variables: random effects of participant ID and fixed effects of seen-status, targeted-user, has-people, age, education, income, geographical region, and the interaction between race and gender.\footnote{\texttt{metric\_val \textasciitilde\ 1 + (1 | participant\_id) + study\_phase + seen + targeted\_user + has\_people + age + education + income + region + (race*gender)}}
These models aligned with the previous findings in Section~\ref{sec:res:seen-status}; the additional variable for study phase only had a significant effect on recognition ($F(1, 8522)=18.58, p < 0.001$).

\begin{figure}[!tb]
  \includegraphics[width=1.0\textwidth]{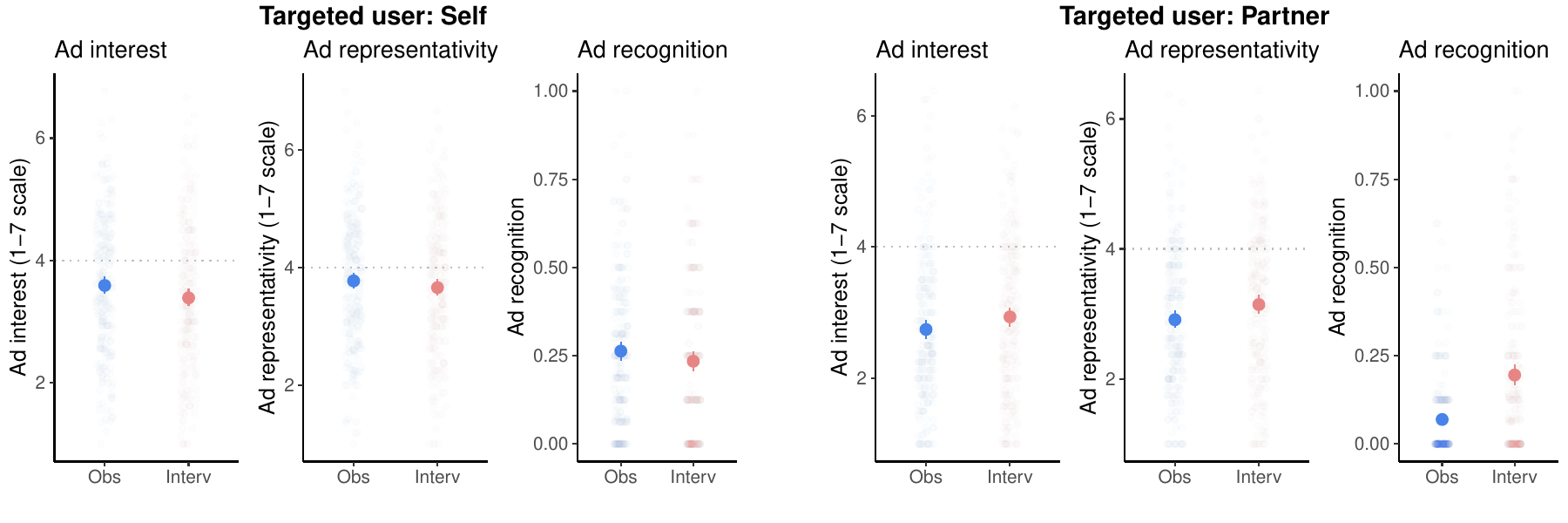}
  \caption{
    Ad metrics by targeted user. (Left) For ads originally targeted to the user (self), sentiment slightly decreased between the observational and intervention phase. (Right) Meanwhile, for ads targeted to their partner, sentiment slightly increased.
  }
  \label{fig:obs_interv_perAd_targetedStatus_change}
\end{figure}

\begin{figure}[!tb]
  \includegraphics[width=0.7\textwidth]{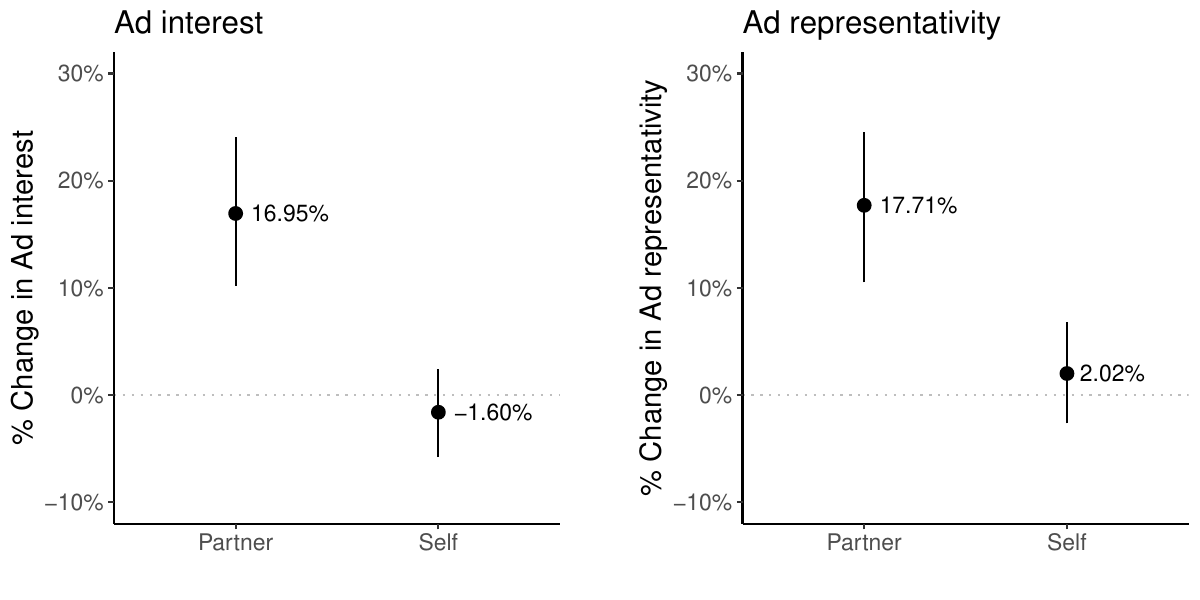}
  \caption{
    Percent change in ad interest and representativity for ads targeted to the user (Self) versus their random swap partner (Partner). We observe notable increases in both metrics for partner ads while metric values for users' own ads remained stable.
  }
  \label{fig:interv_perAd_pctChange_interestRep}
\end{figure}

Moving from these omnibus tests to post-hoc tests examining the direction of effects, we found that the metrics for ads targeted to the random swap partner increased after the intervention, while metrics for ads targeted to the user slightly decreased (Figure~\ref{fig:obs_interv_perAd_targetedStatus_change}). Looking further into this trend, we found that ad interest towards partner ads increased by an average of $16.9\%$ while interest toward one's own ads slightly decreased by an average of $1.6\%$ (Figure~\ref{fig:interv_perAd_pctChange_interestRep}). Similarly, ad representativity for partner ads increased by an average of $17.7\%$ while representativity for one's own ads increased by just $2.0\%$. 
The percentage increases for partner ads correspond to an increase greater than one rating higher on the 7-point Likert scales.
In the intervention phase, users also showed a slight preference for ads that they did not see, and which contained people; see Appendix~\ref{appendix:results-per-ad} for detailed findings on the seen-status and has-people factors.

\subsubsection{Users of marginalized races and genders may benefit more from personalized targeting (RQ2-D)}
\begin{figure}[!tb]
  \includegraphics[width=1.0\textwidth]{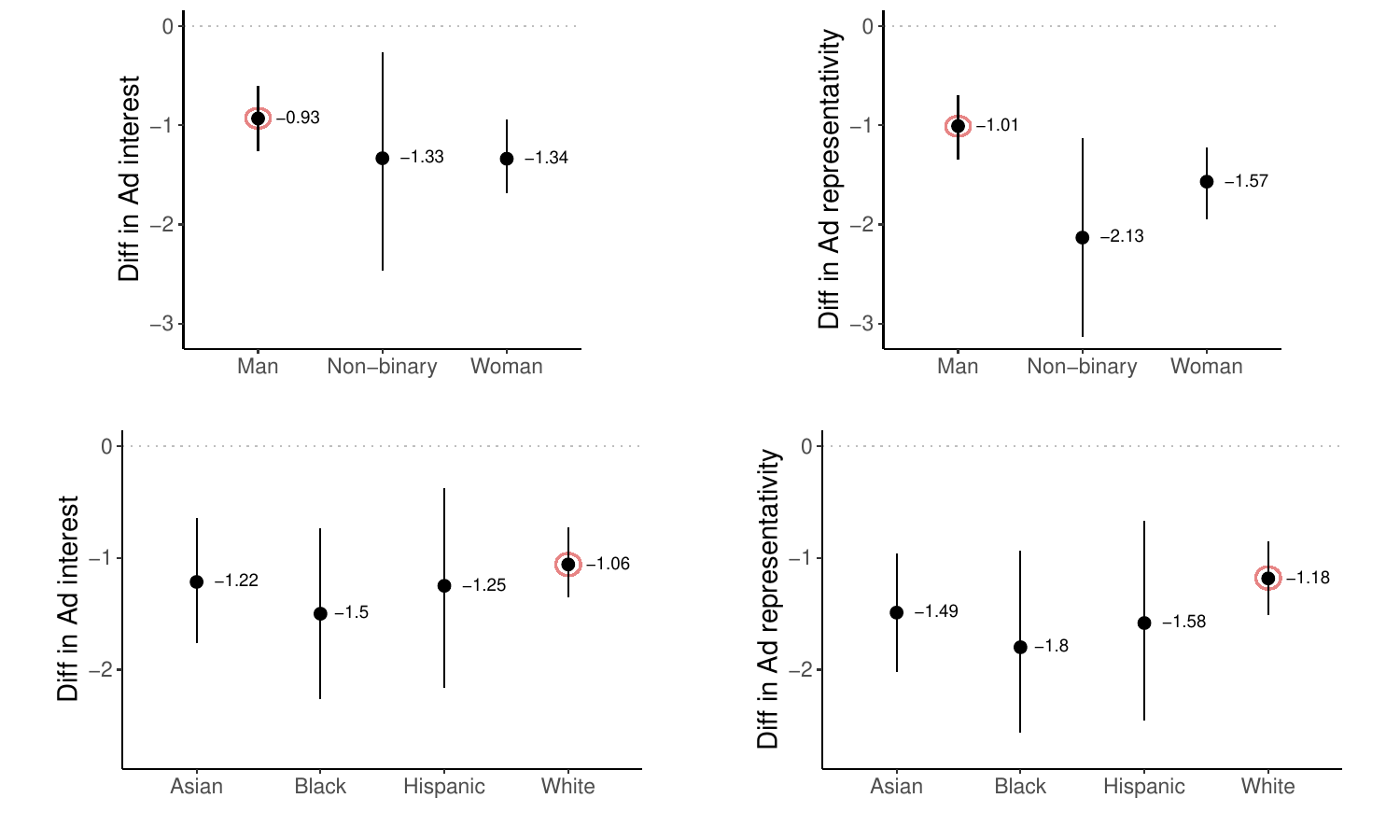}
  \caption{
    Difference in ad interest (left) and ad representativity (right), broken down by participant gender (top) and race (bottom). We observe that the drop in ad metrics in the intervention phase is slightly larger for participants of marginalized races and genders than those who are men or white (annotated in red).
  }
  \label{fig:interv_holistic_genderRace_interestRep}
\end{figure}

While non-targeted ads performed more poorly across the board, we found larger drops in ad performance for participants of marginalized race and gender identities (non-white or non-men).
To investigate the holistic survey results across study phases, we used a linear regression model with a dependent variable of the \textit{difference} in metric value between the intervention and observational phase. We defined $\texttt{metric\_diff} = (\text{intervention value}) - (\text{observational value})$, so a negative value would indicate a drop in the metric value in the intervention phase. The model again had fixed effects of user age, education, income, geographical region, and the interaction between race and gender.\footnote{\texttt{metric\_diff \textasciitilde\ 1 + age + education + income + region + (race*gender)}}
For ad interest, we observed a significant effect of the interaction between race and gender ($F(6, 221)=2.20, p < 0.05$). 
For ad representativity and ad recognition, we did not observe any significant effects among these variables. 
Looking further into race and gender breakdowns for ad interest and representativity, we observed that the drop in metrics appears to be larger for users with genders other than ``man'' and users with races other than ``white'' (Figure~\ref{fig:interv_holistic_genderRace_interestRep}). 

\subsubsection{Users falsely recognize ads targeted to them (RQ2-A)}
\begin{figure}[!tb]
  \includegraphics[width=0.5\textwidth]{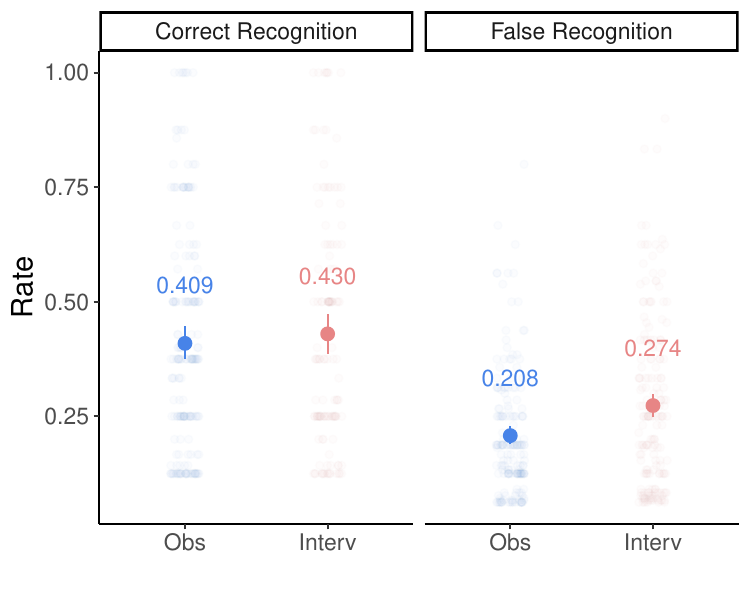}
  \caption{
    Per-ad recognition rates in the observational and intervention phases. We observed comparable \textit{correct}-recognition rates between the phases, but increased \textit{false}-recognition rates in the intervention phase.
  }
  \label{fig:perAd_recall}
\end{figure}

Finally, we observed substantial levels of false recognition in both study phases. 
Already in the observational phase, participants frequently falsely recognized ads at a rate of $20.8\%$ ($SD=13.4\%$).
After the intervention phase, participants' false-recognition rate rose significantly to $27.4\%$ ($SD=18.2\%$); $t(329.5)=-3.88, p < 0.001$, Cohen's $d = 0.41$ (Figure~\ref{fig:perAd_recall}). This rise may have stemmed from participants falsely recognizing ads that were targeted to them, but that they did not in fact see during the intervention phase---such ads may have appeared familiar to participants due to similarity to the \textit{kinds} of ads they are accustomed to seeing. Alternatively, perhaps users may have been exposed to those ads elsewhere, such as on a mobile device or an incognito browser session.

\subsection{Participants and compliance}
\label{section:results-overview}
We now summarize key information on participant retention, demographics, and compliance in our targeted advertising case study.

\subsubsection{Participant retention}
\begin{figure}[!tb]
  \includegraphics[width=0.5\textwidth]{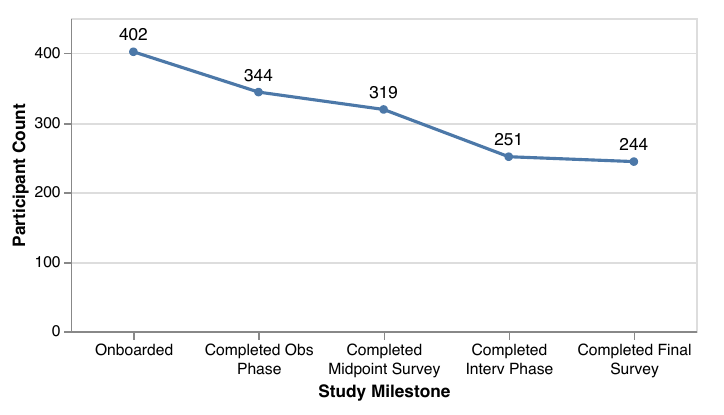}
  \caption{
    Summary of participant drop-off across study milestones.
  }
  \label{fig:participant_dropoff}
\end{figure}

We successfully onboarded 402 participants; of those participants, 244 completed both the observational phase and intervention phase along with the accompanying post-surveys (Figure~\ref{fig:participant_dropoff}). The major drop-off points corresponded to the observational and intervention phases themselves, where some participants did not keep the system active or did not accrue a large enough sample of ads to continue participating; there was an $85.6\%$ retention rate for the observational phase and a $78.7\%$ retention rate for the intervention phase. Of eligible participants, the midpoint survey completion rate was $92.7\%$, and the final survey completion rate was $97.2\%$.

\subsubsection{Participant demographics}
\begin{figure}[!tb]
  \includegraphics[width=1.0\textwidth]{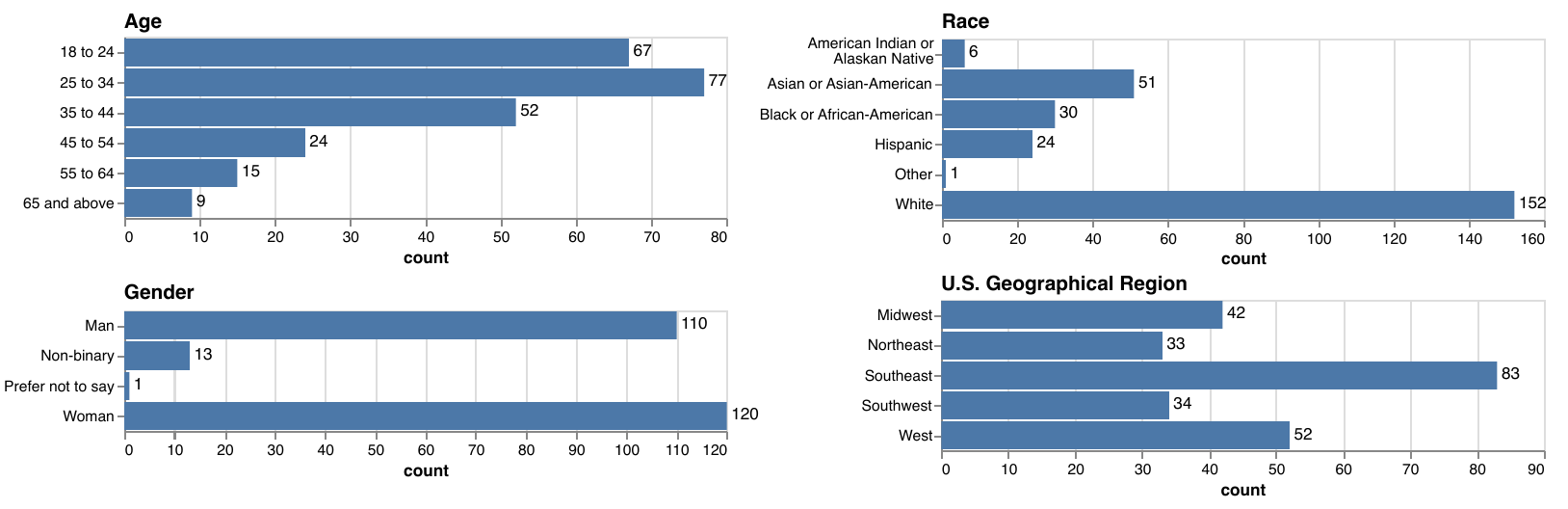}
  \caption{
    Summary of participant demographics ($N=244$) for age, gender, race, and U.S. geographical region.
  }
  \label{fig:dem_overview}
\end{figure}

The final participant pool represented a diverse range of ages, races, and genders (Figure~\ref{fig:dem_overview}). There were 110 men ($45.1\%$), 120 women ($49.2\%$), 13 non-binary participants ($5.3\%$), and 1 participant who preferred not to share their gender. There were 152 participants who identified as white ($62.3\%$), 51 as Asian or Asian-American ($20.9\%$), 30 as Black or African-American ($12.3\%$), 24 as Hispanic ($9.8\%$), 6 as American Indian or Alaskan Native ($2.5\%$), and 1 as Other (percentages here do not sum to $100\%$ since participants could select multiple).

\subsubsection{Participant experience and compliance}
\label{section:compliance}
We found that participants generally had positive study experiences, with an average rating of $5.44$ out of $7$ ($SD=1.31$). 
Participants reported very low non-compliance by disabling the extension ($M=1.29$, $SD=0.74$) or switching to Incognito ($M=1.34$, $SD=0.65$), corresponding to frequency ratings between ``Never (0\% of the time)'' and ``Rarely (<10\% of the time).'' These are self-reported, since we did not track noncompliance. We found that data redaction among our participants was rare: only 62 users (25.4\%) redacted any data, and for those users, on average 4.1\% (median=$0.8\%, SD=7.9\%$) of ads were redacted, or a median of 9.0 redacted ads per user.

\subsection{Advertisement characteristics}
Next, we summarize the volume and nature of advertisements collected during our audit.
We collected $537,945$ advertisements over the course of the study, with $314,762$ advertisements from our final pool of 244 participants. Of those ads, there were $88,604$ observational phase ads, $121,489$ intervention phase original ads, and $104,669$ intervention phase swapped-in ads. 
Overall, $67,402$ ads (21.41\%) were viewed by participants, while the remainder (78.59\%) went unseen.
As expected, ad click-rates were much lower: only $123$ ads (0.04\%) were clicked by participants.
Based on automated person detection, $57,328$ ads we collected (18.21\%) contained at least one person.
Each of these averages across the full data set was also similar when computed on a per-participant basis.
We observed a small set of 5-8 common ad sources (the domain displaying the ad) and targets (the domain to which the ad pointed), and a long tail of sites that occurred far less frequently across users; see Appendix~\ref{appendix:ad-sources-targets} for details.
A preliminary qualitative analysis of ad image content found a variety of product categories (e.g., hardware, furniture, clothing, food, athletic gear, tech products, financial services, employment, and educational opportunities), image contents (e.g., celebrities, brand logos, stock photos), and ad formats (e.g., horizontal banner ads, product images, animated gifs, and discount offers); see Appendix~\ref{section:ad-analyses} for details.

\subsection{Data verification}
We also performed a post-study analysis to benchmark Intervenr's performance in detecting real-life participants' ads and the person detection model's performance on advertising images.

\subsubsection{Ad detection coverage}
To validate the coverage of Intervenr's ad detection in practice, we gathered a random sample of 150 URLs among the set of URLs visited by study participants. Then, for each sampled URL, a member of the research team visited the site and manually counted the number of ads that appeared on the page as a ground truth value against which to compare the number of ads collected by Intervenr. Based on bootstrap sampling with 500 resamples of our dataset, we observed a median ad coverage rate of $80\%$ (the proportion of manually-detected ads that were captured by Intervenr), and we observed a false negative rate of 15.3\% (the rate at which Intervenr did not detect ads that were manually detected).
We conclude that Intervenr captures a relatively high proportion of the ads that users encounter across the web.

\subsubsection{Person detection coverage}
We also validated the coverage of our automated person detection model on the images captured in our study, given that the images may have differed from the model's COCO training dataset. We gathered a random sample of 350 ad images from the observational phase of the study, and a member of the research team viewed each sampled image and manually annotated the count of people contained in the image. We then compared these ground truth labels with the person detection model's results. In this 350-image dataset, only 81 images (23.1\%) contained a person based on manual annotation, which was comparable to the classifier-estimated proportion of ads that contained people (18.21\%) for the full participant dataset with over 314k ads. 
Based on bootstrap sampling with 500 resamples on this initial dataset, we observed an accuracy of $86.3\%$ ($\pm 3.7\%$) and a false negative rate of $10.5\%$ ($\pm 3.7\%$).

\section{Discussion}
\label{sec:discussion}
Having introduced sociotechnical audits, the Intervenr system to enact such audits, and a case study on targeted advertising, we highlight key takeaways and insights alongside limitations and future work for each of the three components.

\subsection{Case Study: Targeted Advertising}
We begin with the most specific contribution: our case study. Our audit of targeted advertising provides insights on both sides of the sociotechnical system. Overall, we found that targeted ads perform only moderately well; in terms of interest and representativity, users have low-to-moderate sentiment towards the ads they see in-browser. 
Participants only correctly recognized ads that they had seen about 41\% of the time, and false recognition (claiming to have seen an ad they did not) was around 21\%. We also audited users' interactions with this sociotechnical system using our ablation intervention. In doing so, we were able to confirm that personalized ads performed better across all metrics---including recognition, interest, and representativity---compared with ads targeted to someone else. However, after only a week of this intervention, the random ads' performance improved as users appeared to be acclimating in a very short amount of time.  

Enabled by our sociotechnical lens, this study design suggests that although personalized ad targeting is premised upon the importance of being responsive to user behaviors (so much so that this drives a multi-billion dollar industry), users are also changing in response to their algorithmic environment. Their positive responses to ad targeting may be inflated by the positive effects of familiarity and repeated exposure. This dual perspective calls into question the actual need for invasive surveillance of users and sale of their personal data that underpins the ad industry. 

As with most auditors, we are also interested in understanding the experiences of marginalized users and the potential for inequity caused by technological systems. In this study, we found that breaking ad targeting had some more negative impacts on participants with marginalized race and gender identities, in comparison to white and/or male participants. This does raise the important point that more personalized content may hold greater value for those whose identities place them outside the social ``default,'' and that simply removing all targeting may be contraindicated for some categories of users. 

\subsubsection{Limitations}
Although (or perhaps because) our study is the first sociotechnical audit of targeted advertising to our knowledge, it has many limitations; here, we aim to broach several of the study's major limitations.
First, since our audit only captures browser content (a limitation of the Intervenr tool that we discuss further in \ref{sub:discussion-intervenr-limits}), we know we have only partial coverage of users' ads ecosystem. This has several implications for measures including biasing recognition (which could be influenced by ads users saw elsewhere), and weakening the effect of our intervention (since users almost certainly saw their own targeted ads on their other devices). This does mean readers should take measures like recognition with a grain of salt, but also that our finding that users acclimate to non-targeted ads is likely an underestimate: this effect might be stronger if the intervention had more complete coverage. 
Next, while our study was longer than most prior work in this space, it only spanned two weeks, whereas we know that users' interactions with this sociotechnical system are happening constantly over the course of their entire lives. We cannot speak to the effects that a longer study might find, as people's acclimation to other ads may at some point plateau or reverse. Evaluating algorithmic impact on people takes longer than evaluating just the algorithm, and we encourage future sociotechnical auditors to consider a longer time-horizon (four weeks or even longer) when auditing the human aspect of a system. 
Finally, we also cannot fully account for non-compliance, as participants could use a different browser, incognito window, or another device. We measured self-reported non-compliance as described in Section~\ref{section:compliance} and find it to be low, but participants may have felt pressure to underestimate their non-compliance. We encourage these limitations to be carefully considered and mitigated when possible in future work. 

\subsubsection{Future Work} Given the findings of our case study, we encourage further research that speaks to the level of surveillance and privacy invasion that targeted advertising currently requires, and whether its material benefits justify those costs. Future sociotechnical auditors should evaluate user responses to other alternatives, including completely random sets of ads, ads they have some hand in curating themselves, or no ads at all. These evaluations should be carried out with an attention to users' own positionality and psychological qualities like belongingness, as we find evidence that proximity to social power shapes people's experiences with sociotechnical systems. Any work to improve these systems must be done while centering the needs of users who need it the most. 

There is also considerable room for sociotechnical audits on advertising to take this work further, for instance by empowering users to take a more active role in the audit process, as other audit methods have demonstrated~\cite{lam2022end}. Future work might use qualitative methods to understand participant experiences or conduct deeper analyses of ads imagery. Prior work suggests that there should be significant qualitative differences between the ads shown to different users based on the way these algorithms are known to function~\cite{sapiezynski2022algorithms}; we have only scratched the surface with our own efforts, in Appendix~\ref{section:ad-analyses}, to qualitatively and computationally understand the content of ad images. 

\subsection{Intervenr System}
Zooming out one level, we discuss our second contribution, Intervenr. While less critical to our specific study, and therefore highlighted less in this paper, an important characteristic of Intervenr is its reusability. The system was designed to be a flexible asset that might allow researchers to semi-automatically conduct a wide range of longitudinal, in-browser sociotechnical audits. Adapting Intervenr to study targeted ads required ads-specific modifications (outlined in Appendix \ref{appendix:ad-infra-details}), but once set up, the system ran our audit with a high degree of automation. It was relatively straightforward to scale the system up to simultaneously perform in situ ad-swapping interventions on hundreds of users, with computing capacity to process thousands of ad images nightly.

\subsubsection{Limitations and Future Work}
\label{sub:discussion-intervenr-limits}
While effective and powerful as a tool, Intervenr does have several limitations that can inform future design directions for sociotechnical auditing infrastructure. First, the system required considerable ads-specific customizations that would not be needed in future studies in other domains, predominantly the integration of AdNauseam for ads tracking. Building general-purpose infrastructure is challenging, and the range of potential auditing domains is very broad, so future tools may want to focus on a more specific type of sociotechnical system to audit rather than attempting to support to all browser-based systems. 

Second, maintenance of such a tool is an ongoing challenge---at the time of writing, Google has announced changes to Chrome (the move to Manifest V3 from V2) that will limit the capabilities of web extensions like ours. These changes may restrict our ability to enact the in-browser interventions that are critical in this work, especially those related to ad blocking. 
However, working with the Mozilla Rally infrastructure, which is capable of executing sociotechnical audits in Firefox, is a promising direction for future work~\cite{mozilla2021rally}. Even in the absence of programmatic access, other researchers have used methods like screen-capturing to collect data~\cite{reeves2021screenomics}.

Finally, and most glaringly, our system is desktop-only, while research has found that approximately 15\% of American adults are smartphone-only users~\cite{pew2021mobile}. Desktop data is still relevant since most smartphone users are equally or more likely to use a desktop computer~\cite{gallup2015smartphone}, but this is a major limitation of most research in this space, since technical approaches for collecting in-app mobile data are very limited. This is a major area for future work supporting sociotechnical audits and traditional algorithm audits.

We hope that Intervenr can be used to run additional sociotechnical audits in the future, including by researchers with less technical expertise than would be required to build such a system entirely from scratch. Since Intervenr has the capacity to collect so much sensitive user data, we are in the process of exploring ways to make the system accessible to other researchers without making the tool fully open-source. 

\subsection{Sociotechnical Auditing}
Our headlining contribution in this work, which motivates our system and case study, is the concept of a sociotechnical audit. We do not consider our paper to have contributed the first sociotechnical audit. For instance, as referenced in Section \ref{sec:relwork}, there are several notable examples where auditors have paired traditional algorithm audits with user audits, for example in the form of in-browser or on-platform interventions~\cite{mozilla2021rally,matias2023influencing}, or with separate controlled experiments~\cite{metaxa2021image}.

\subsubsection{Limitations and Future Work}
We see several opportunities to extend the sociotechnical audit approach we have presented. One major barrier of this approach is the need to work directly with users and user-facing systems, which can be challenging and costly. Research strategies like algorithm audits and RCTs have developed many workarounds to this issue (e.g., sock puppets or crowdworkers). In kind, sociotechnical auditors will need to creatively iterate on their study designs and implementations to address the effort barriers and financial barriers of the method. However, existing evaluation methods may lend inspiration. For example, given the challenges of limited third-party access, future auditors might develop hybrid approaches that involve partial access to a system and its users, along the lines of work on first- and second-party algorithm audits~\cite{raji_internal_auditing, wilson2021pymetrics}. One or more platforms might agree to take part in such an audit by surfacing a limited API for auditors to implement their study, or they might execute an auditor-designed study and provide auditors the data to analyze. Such an approach might be a convenient way for auditors to benefit from expanded algorithm intervention control and free users while not requiring full internal access. However, we note that such hybrid arrangements are still contentious in the algorithm auditing community, since they may be seen as jeopardizing the impartiality of auditors. 

Enhancing participants' agency in the context of sociotechnical audits is another major direction for future work. Recent algorithm auditing research has noted that audits are typically conducted by a relatively small group of technical experts who have the necessary skills to design, implement, and report an audit~\cite{costanza_chock_who_audits_the_auditors}. However, everyday users have a great deal of expertise about their own communities and how algorithms impact them~\cite{shen_everyday_alg_audit}. 
As with algorithm auditing, sociotechnical auditing would benefit from tools that empower end users to lead their own audits and study harmful algorithmic behavior that auditors miss~\cite{lam2022end}.

Lastly, an exciting byproduct of this method is its capacity to envision and test better alternatives to our current sociotechnical systems. This is also a major challenge: developing and deploying compelling and well-executed interventions is no easy task. In some cases, we may opt to simply test users with algorithmic behavior that already exists in a current or historical system (e.g., comparing a social media site's ranked feed with a chronological version). However, our STA method beckons us to imagine bolder, more creative, or more socially-equitable alternatives. Building on algorithm audits' evaluation of the present, sociotechnical audits can help us imagine the future.

\section{Conclusion}
Algorithm auditing serves as a critical accountability method to understand black-box technical systems and hold them to legal or ethical standards. But many systems, especially those of interest to the CSCW community, are \textit{sociotechnical}---their social, human aspects are as essential as any technical components. For such systems, algorithm audits only capture one part of the story. In this paper, we introduced the concept of \textit{sociotechnical auditing}, a methodology for auditing algorithmic systems through a wider, sociotechnical lens with the goal of capturing how users and algorithms jointly influence each other. 
To demonstrate our method, we developed the \textit{Intervenr} system, which enables auditors to enact two-phase, longitudinal, browser-based sociotechnical audits with consenting, compensated participants. This two-phase design, comprised of an audit of the system's algorithmic aspects in the first phase, and of its social aspects (such as its users' changing behaviors over time) in the second, is one possible form sociotechnical audits can take.
Finally, we conducted a two-week sociotechnical audit of targeted online advertising ($N=244$), first auditing the current state of ad targeting, and subsequently deploying an ablation intervention to audit users' relationship to ad targeting systems. Our case study finds that personalized targeted ads indeed perform better with users, but that users appear to acclimate to a random partner's ads after only a week of exposure. The sociotechnical auditing method allows us to understand the interplay between ad targeting and its users, suggesting that targeted ads' performance may be driven in part by repeated exposures.
These results question whether targeting advertising indeed maximizes benefit for users or whether alternatives might fare comparably for them, especially given the cost of practices that invasively surveil and capitalize on user data.
While algorithm audits help us to hold today's \textit{existing} technical systems accountable for their behavior, sociotechnical audits help us to develop a broader understanding of how technical and human components intertwine, allowing us to envision and experiment with \textit{alternative algorithm designs} beyond the status quo.

\begin{acks}
We thank our anonymous reviewers; our Ph.D. advisors Michael Bernstein, James Landay, and Jeffrey Hancock; and colleagues Joon Sung Park and Helena Vasconcelos for their valuable feedback on our paper. We thank Daniel Howe for sharing helpful guidance on the AdNauseam system, and we thank Megan Mou, Boyang Jia, and Kyrhee Powell for their assistance with data analysis. This project was funded by the Brown Institute for Media Innovation at the Stanford School of Engineering and by the Stanford Institute for Human-Centered Artificial Intelligence (HAI).
\end{acks}


\received{January 2023}
\received[revised]{April 2023}
\received[accepted]{May 2023}

\bibliographystyle{ACM-Reference-Format}
\bibliography{biblio}

\appendix
\section{Appendix}
\subsection{Intervenr Implementation Details}
\label{appendix:implem-details}
\subsubsection{Browser extension}
The browser extension is implemented as a Google Chrome extension using Manifest Version 2 with two independent content scripts. The \textit{registration} script handles the linking between the participant's extension and their account on the web application. The \textit{reconnect} script performs this linking again at the user's request (from the web app homepage) if the link has been broken. Additionally, one or more \textit{custom content scripts} are responsible for handling the main intervention-related logic to (1)~collect data from the webpage and send it the web app and/or (2)~receive data from the web app to modify contents of the webpage. These can be modified to perform the relevant intervention logic for a particular sociotechnical audit.

\subsubsection{Web application}
Our web application is a Django web app with a PostgreSQL database. In production, we run the web app on a Heroku 1X web server and with an Amazon RDS PostGreSQL database.
The Django backend consists of the following components: (1)~\textit{onboard}: classes supporting the user onboarding flow and forms, (2)~\textit{frontend}: classes supporting all of the participant-facing and auditor-facing interfaces, including surveys, and (3)~\textit{extension}: classes supporting communication with the browser extension and intervention-related database reading and writing.

\subsubsection{Data analysis pipeline}
The data analysis pipeline is set up on an Amazon EC2 server with several custom Python scripts that are configured to run using cron jobs. These scripts can be customized depending on the particular sociotechnical audit, but examples include scripts to download persistent copies of images to a Amazon S3 bucket, scripts to resolve and store URLs, and scripts to perform natural language processing and computer vision tasks on raw text or images. These scripts commonly read from and write to the RDS database, enabling subsequent downstream analysis via queries to the RDS database in PostgreSQL applications or computational notebooks.

\subsection{Ad-specific Infrastructure Details}
\label{appendix:ad-infra-details}
\subsubsection{Ads-specific browser extension}
Our extension needed to detect and collect ads, swap in new ad content, and track user views and clicks on ads.
Our extension built on AdNauseam\footnote{\url{https://adnauseam.io}}, an open-source browser extension that itself builds on top of uBlock Origin\footnote{\url{https://github.com/gorhill/uBlock}}, a content-blocking extension that blocks ads, trackers, and malware sites. We selected AdNauseam because it both incorporates up-to-date ad detection as well as logic to extract and save detected ads~\cite{nissenbaum2009trackmenot}.
Starting from this extension, we integrated the Intervenr browser extension functionality that communicates with our web application. We then added functionality to send detected ads to our web app to be stored in the database by sending requests to web app endpoints.
Then, a large portion of our additional implementation work involved the functionality to perform ad swapping. This extension logic was responsible for dynamically removing existing ads from the webpage, querying our web app for swap-ads, and altering the participant webpage code to insert swap-ads in the same positions as the old ads. 

Finally, we added custom user view and click tracking. We implemented ad view tracking using the Intersection Observer API\footnote{\url{https://developer.mozilla.org/en-US/docs/Web/API/Intersection_Observer_API}}, which allowed us to compare target elements (webpage elements containing ads, in our case) with the user's visible viewport. This allowed us to track whether each of the ads that are delivered to the user on a webpage actually entered the visible portion of the screen. We added this view-tracking both to original ads in the observational phase and to swap-ads in the intervention phase. Then, we implemented ad click tracking using JavaScript event listeners for clicks on ad webpage elements. We again added this click-tracking both to original ads in the observational phase and to swap-ads in the intervention phase.

\subsubsection{Web application}
We extended our web application to communicate with our ads-specific browser extension to enable our intervention, and we set up the custom sampling and user interfaces required for our surveys.
These changes included implementing endpoints to respond to several request types: storing detected ads, updating ad view and click counts, and fetching swap-ads to render on participant webpages. We also added functionality to automatically alter the system's behavior based on the study phase. During the observational phase, the endpoints only recorded original ads. In the intervention phase, they recorded both original ads and swap ads. 

We also extended the auditor admin functionality to allow us to assign random swap partners and preview participant surveys, and added custom dashboard views to track study metrics including ads collected and survey completion for all participants.

To instantiate the survey design described in Section~\ref{section:survey-design}, we created custom Django Forms in the backend to support our questions and set up frontend interface pages to render them. The backend implementation included logic to randomly sample holistic and per-ad questions. To reduce load-time for users, we pre-generated and cached these ad samples before surveys were released to participants.

\subsubsection{Data analysis pipeline}
\label{section:ad-data-analysis-pipeline}
We extended our data analysis pipeline to support ads-specific data processing needs. First, we needed to resolve the links associated with ads, since oftentimes the links used on ads are shortlinks that redirect (often multiple times) to different final sites. This link-resolving process can take longer than a typical page load time and can disrupt participant experiences if conducted on the fly. We created a link-resolving script that ran nightly on our data analysis server, iterating through all ads collected that day and resolving their links.

Next, for our survey to include ads with and without people, we performed post-processing on all images to determine whether they contained people. We used \texttt{cvlib}\footnote{\url{https://www.cvlib.net/}}, an open-source computer vision library for Python, to perform object and person detection. This library uses the YOLOv3 model trained on the COCO dataset, which features 80 common object categories.\footnote{\url{https://github.com/arunponnusamy/object-detection-opencv/blob/master/yolov3.txt}}
The object-detection script again ran nightly on our data analysis server, iterating through all image ads from the past day, running images through the object-detection model, and storing the results. 
Since this processing is more computationally expensive, we distributed this processing across multiple servers. 

Lastly, advertisement images were extracted as URLs to remotely-hosted images, but these URLs often go stale over time. To avoid losing access to ad images, we downloaded all ad images to persistent storage using another nightly script.

\subsection{Results: Per-ad survey results for seen-status and has-people}
\label{appendix:results-per-ad}
\begin{figure}[!tb]
  \includegraphics[width=0.9\textwidth]{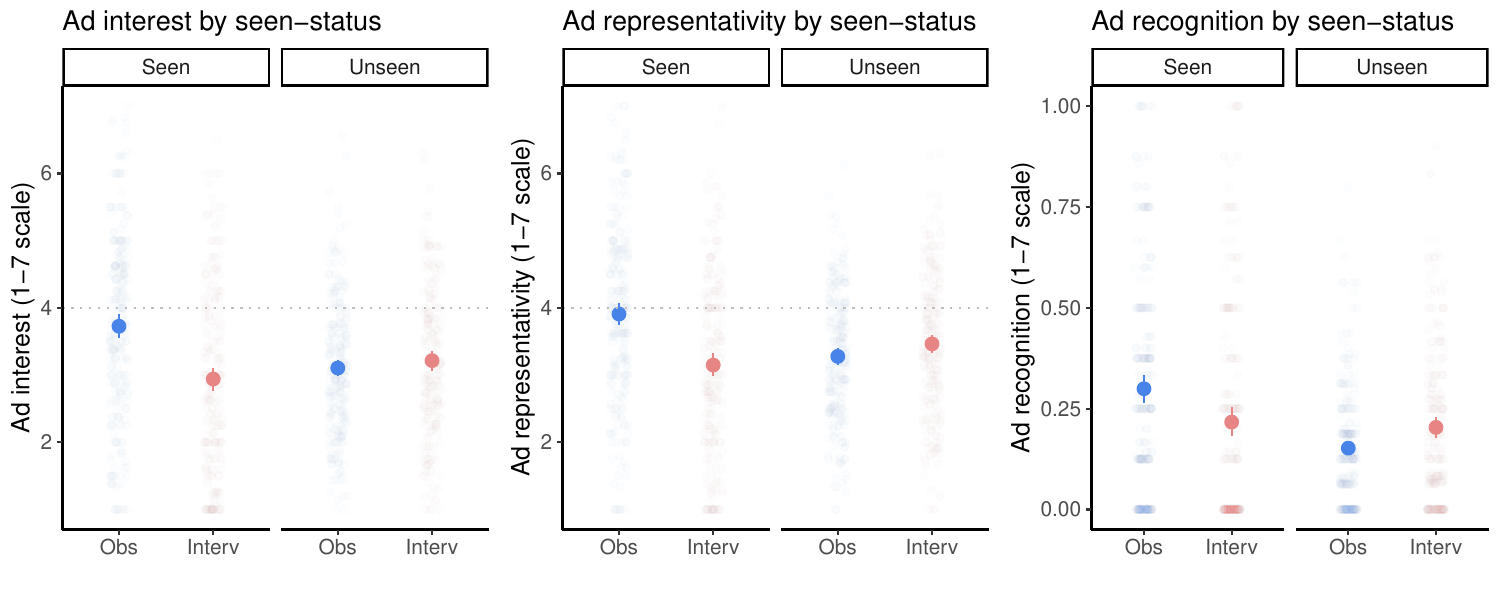}
  \caption{
    Ad metrics by seen-status. In the observational phase, we observe higher metric values for ads that were seen (participants' own ads), but in the intervention phase, ads that were seen and unseen have more comparable metric values.
  }
  \label{fig:obs_interv_perAd_seenStatus}
\end{figure}

\begin{figure}[!tb]
  \includegraphics[width=0.9\textwidth]{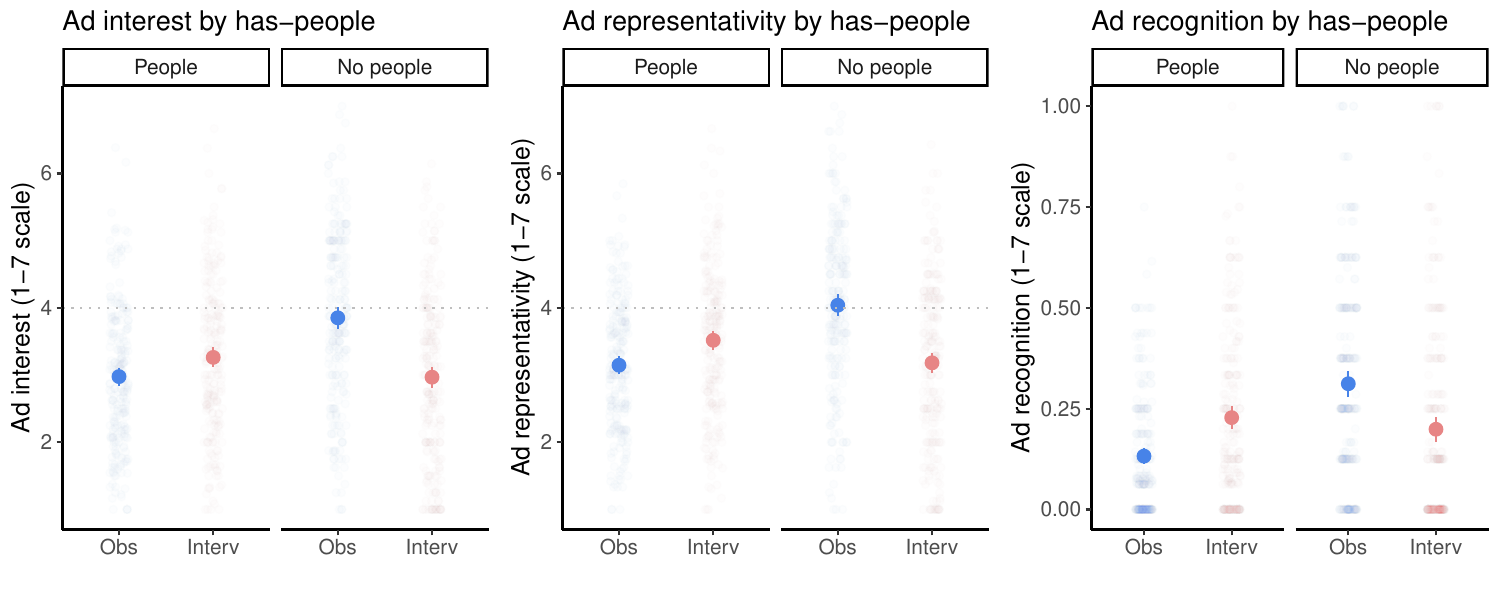}
  \caption{
    Ad metrics by has-people. In the observational phase, ads without people achieved higher metric values, but in the intervention phase, ads with people received slightly higher metric values.
  }
  \label{fig:obs_interv_perAd_pplStatus}
\end{figure}

For the factor of seen-status, we observed that ad interest, representativity, and recognition all decreased for ads that were seen and maintained roughly stable for ads that were not seen (Figure~\ref{fig:obs_interv_perAd_seenStatus}).
Then, for the factor of has-people, we observed that the three metrics decreased for ads that did not contain people and increased for ads that did contain people (Figure~\ref{fig:obs_interv_perAd_pplStatus}).

\begin{figure}[!tb]
  \includegraphics[width=0.9\textwidth]{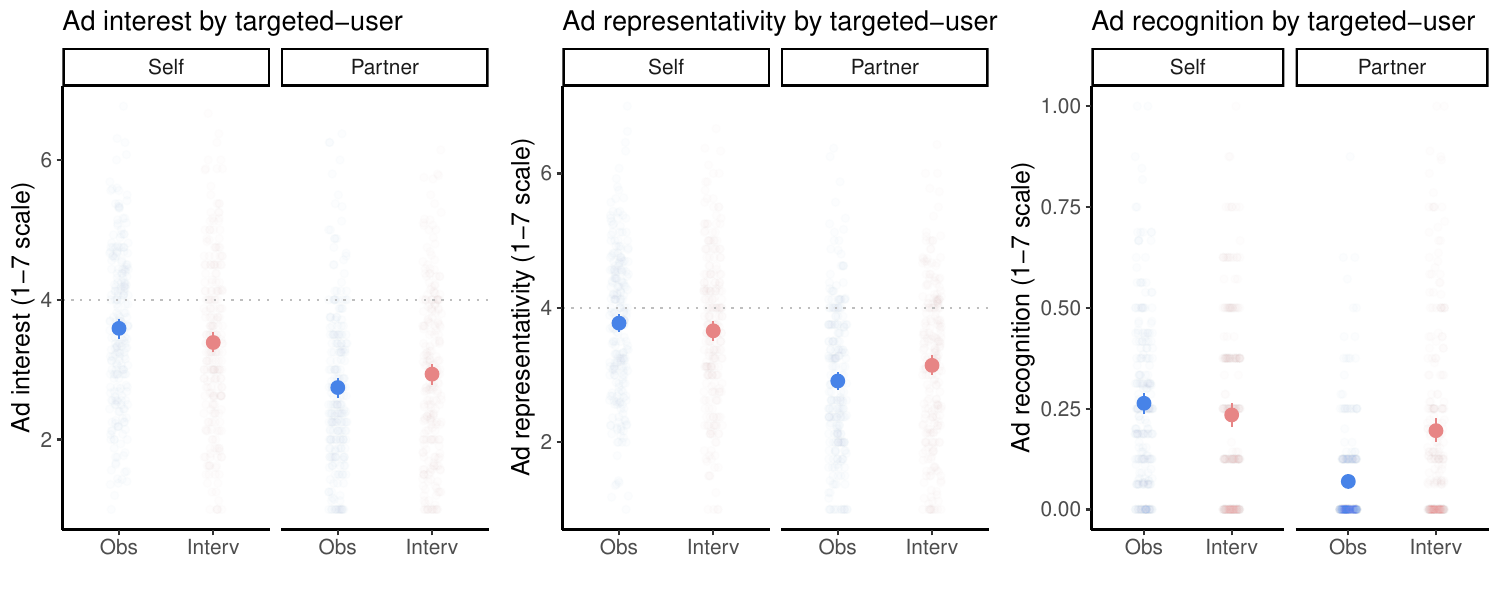}
  \caption{
    Ad metrics by targeted-user. In the observational phase, we observe higher metric values for ads targeted to the user (comparing the two blue points in each chart), but in the intervention phase, sentiment towards partner ads increases (comparing the blue and red points in the two rightmost columns in each figure).
  }
  \label{fig:obs_interv_perAd_targetedStatus}
\end{figure}

\subsection{Results: Ad Sources and Targets}
\label{appendix:ad-sources-targets}

\textit{Ad sources}. Out of all ads collected, the most common ad sources---the domain on which the ad appeared---were Reddit (23.4\%), Yahoo (15.46\%), Google (12.76\%), Facebook (3.8\%), and Ebay (2.17\%), with other sources comprising 38.82\% of ads.
Since these proportions can be impacted by the volume of ads that different sites display on the page, we also looked at the proportion of participants whose ads included each ad source. We found that across participants, the most common ad sources were Google (6.14\%), Reddit (2.54\%), Amazon (1.75\%), and Yahoo (1.48\%).

\textit{Ad targets}. Out of the full ad pool, the most common ad targets---the domain an ad linked to---were Google (43.34\%), Reddit (11.89\%), Yahoo (8.0\%), Amazon (4.07\%), LinkedIn (2.02\%), Facebook (1.51\%), Walmart (1.18\%), and Dianomi (1.13\%), with other targets comprising 26.87\% of ads.
Breaking down by the proportion of participants whose ads included each ad target, we observed that the most common ad targets were Google (6.26\%), Amazon (1.56\%), Reddit (1.42\%), Yahoo (1.29\%), and Walmart (1.19\%).
These ad target links are often internal links that redirect to other external pages, so we also looked into the resolved target links. We found that out of all ads, the most common resolved at target links were Reddit (11.98\%), Yahoo (6.92\%), Google (5.88\%), Amazon (4.57\%), Walmart (4.11\%), Facebook (1.69\%), and Capital One Shopping (1.60\%).

\subsection{Results: Additional Advertisement Analyses}
\label{section:ad-analyses}
We conducted preliminary qualitative analyses to provide additional context on the content of ads delivered to our study participants, relying on automated methods such as image clustering and automated image classification to give us a rough idea of high-level trends in ad image content.

\subsubsection{Qualitative analyses}
Given the scale of our ad collection with over 314k ads and over 88.6k ads in the observational phase, we performed clustering on a random sample of ad images and qualitatively analyzed the resulting clusters. We filtered to observational phase ads and selected a random sample of 10,000 ads. Then, we used CLIP, a pretrained image-to-text model, to transform raw images to 512-dimensional embeddings~\cite{CLIP}. We next applied UMAP (uniform manifold approximation and projection) to perform additional dimensionality reduction~\cite{2018arXivUMAP}. Finally, we passed these reduced embeddings to the HDBSCAN (hierarchical density-based spatial clustering of applications with noise) clustering algorithm~\cite{mcinnes2017hdbscan}. Since HDBSCAN is density-based, it does not require upfront specification of fine-grained parameters such as the number of total clusters, and it can tolerate noisy data.

This method assigned 1,976 images to sixteen clusters of size 20 or greater, while the remaining images were not assigned to a cluster of sufficient size. For each cluster, a member of the research team reviewed the images and summarized high-level themes. The cluster summaries and respective frequency (out of images assigned to clusters) are shown in Table~\ref{table:cluster_summary}. We find a broad variety of ad types in terms of \textit{product categories} (e.g., hardware, furniture, clothing, food, athletic gear, tech products, financial services, employment and educational opportunities), \textit{image contents} (e.g., celebrities, brand logos, stock photos), and \textit{ad formats} (e.g., horizontal banner ads, product images, animated gifs, coupons and offers). 

\begin{table*}[!tb]
  \centering
  \scriptsize
    \begin{tabular}{p{0.4\textwidth} p{0.05\textwidth} p{0.1\textwidth}}
    \toprule
    \textbf{Ad Cluster Summary} & \textbf{Count} & \textbf{Percentage}\\
    \midrule
    \textbf{Bottled products} & 
    {453} &
    {22.9\%}\\[0.1cm]

    \textbf{Celebrity headshots} & 
    {289} &
    {14.6\%}\\[0.1cm]

    \textbf{Horizontal banner ads} & 
    {257} &
    {13.0\%}\\[0.1cm]

    \textbf{Hardware and equipment} & 
    {236} &
    {11.9\%}\\[0.1cm]

    \textbf{Coupons, discounts, and offers} & 
    {104} &
    {5.3\%}\\[0.1cm]

    \textbf{Stock photos of offices, work, and collaboration} & 
    {102} &
    {5.2\%}\\[0.1cm]

    \textbf{AdChoices logo} & 
    {101} &
    {5.1\%}\\[0.1cm]

    \textbf{Natural food and ingredients} & 
    {96} &
    {4.9\%}\\[0.1cm]

    \textbf{Black-and-white photos} & 
    {75} &
    {3.8\%}\\[0.1cm]

    \textbf{Sneakers and athletic goods} & 
    {61} &
    {3.1\%}\\[0.1cm]

    \textbf{Fashion, clothing, and shoes} & 
    {51} &
    {2.6\%}\\[0.1cm]

    \textbf{Furniture and home goods} & 
    {43} &
    {2.2\%}\\[0.1cm]

    \textbf{Animated horizontal banner GIF ads} & 
    {31} &
    {1.6\%}\\[0.1cm]

    \textbf{Boxed and bagged products} & 
    {29} &
    {1.5\%}\\[0.1cm]

    \textbf{Technology and product brand-oriented ads} & 
    {24} &
    {1.2\%}\\[0.1cm]

    \textbf{CapitalOne promo code ad} & 
    {24} &
    {1.2\%}\\
    \bottomrule
    \end{tabular}
    \caption{
        Summary of ad image clusters, sorted by frequency. The clusters span a variety of product categories, image contents, and ad formats.
    }
    \label{table:cluster_summary}
\end{table*}

\subsubsection{Object analyses}
\begin{figure}[!tb]
  \includegraphics[width=0.7\textwidth]{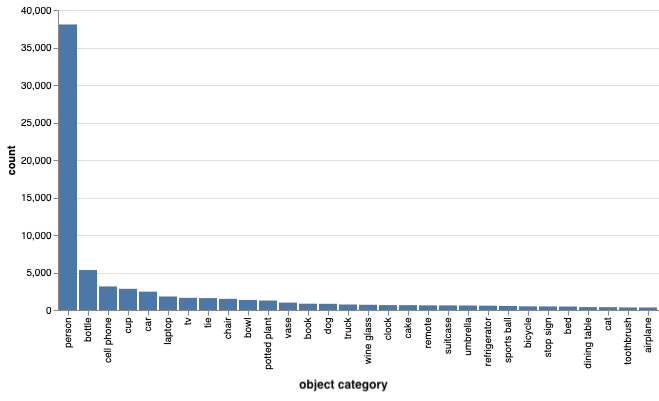}
  \caption{
    Frequency of object categories among observational phase ads.
  }
  \label{fig:obj_frequency}
\end{figure}

We performed automated object detection to gather high-level trends in the objects contained in participant ad images.
As described in Section~\ref{section:ad-data-analysis-pipeline}, all ad images were processed with a computer vision model for object detection, which captures 80 object categories, including a ``person'' category.
We found that among observational phase ads, 52.5\% contained any of the object categories, and 27.3\% contained people. Figure ~\ref{fig:obj_frequency} reports those occuring most frequently.
For comparison, people were detected more frequently than any object---around 7 times the number of occurrences of the second most common object category, ``bottle.'' 

\end{document}